\documentclass[%
prx,
 amsmath,amssymb,
 aps,
twocolumn,
longbibliography,
showkeys
]{revtex4-1}

\usepackage{graphicx}
\usepackage{epstopdf}
\usepackage{dcolumn}% Align table columns on decimal point
\usepackage{bm}% bold math
\usepackage{color}
\usepackage[mathlines]{lineno}
\usepackage[breaklinks=true,colorlinks=true,linkcolor=blue,urlcolor=blue,citecolor=blue]{hyperref}
\usepackage{ragged2e}
\usepackage{enumitem}
\usepackage{bibunits}
\usepackage{amsthm}
\usepackage{diagbox}

\defaultbibliographystyle{elsarticle-num}
\defaultbibliography{biblio}

\renewcommand\thefigure{\arabic{figure}}

\begin{document}

%\preprint{APS/123-QED}

\title[]{Simple bots breed social punishment in humans}% Force line breaks with \\

\author{Chen Shen$^1$}
%\email{steven\_shen91@hotmail.com}
\author{Zhixue He$^2$}
\author{Lei Shi$^2$}
\email{shi\_lei65@hotmail.com}
\author{Zhen Wang$^3$}
\email{w-zhen@nwpu.edu.cn}
\author{Jun Tanimoto$^{4,1}$}
\email{tanimoto@cm.kyushu-u.ac.jp}

\affiliation{
\vspace{2mm}
\mbox{1. Faculty of Engineering Sciences, Kyushu University, Kasuga-koen, Kasuga-shi, Fukuoka 816-8580, Japan}
\mbox{2. School of Statistics and Mathematics, Yunnan University of Finance and Economics, Kunming 650221, China}
\mbox{3. Center for OPTical IMagery Analysis and Learning (OPTIMAL) and School of Mechanical Engineering,}
\mbox{Northwestern Polytechnical University, Xi'an 710072, China}
\mbox{4. Interdisciplinary Graduate School of Engineering Sciences, Kyushu University, Fukuoka, 816-8580, Japan}
}

\date{\today}
% It is always \today, today,
% but any date may be explicitly specified

\begin{abstract}
Costly punishment has been suggested as a key mechanism for stabilizing cooperation in one-shot games. However, recent studies have revealed that the effectiveness of costly punishment can be diminished by second-order free riders (i.e., cooperators who never punish defectors) and antisocial punishers (i.e., defectors who punish cooperators). In a two-stage prisoner's dilemma game, players not only need to choose between cooperation and defection in the first stage, but also need to decide whether to punish their opponent in the second stage. Here, we extend the theory of punishment in one-shot games by introducing simple bots, who consistently choose prosocial punishment and do not change their actions over time. We find that this simple extension of the game allows prosocial punishment to dominate in well-mixed and networked populations, and that the minimum fraction of bots required for the dominance of prosocial punishment monotonically increases with increasing dilemma strength. Furthermore, if humans possess a  learning bias toward a "copy the majority" rule or if bots are present at higher degree nodes in scale-free networks, the fully dominance of prosocial punishment is still possible at a high dilemma strength. These results indicate that introducing bots can be a  significant factor in establishing prosocial punishment. We therefore, provide a novel explanation for the evolution of prosocial punishment, and note that the contrasting results that emerge from the introduction of different types of bots also imply that the design of the bots matters.   
\end{abstract}


\maketitle

% \begin{bibunit}

 \section{Introduction}
In human societies, people are willing to help genetically unrelated strangers even at expense of their own interest, however, this behavioral pattern cannot be explained by the principle of the "survival of the fittest"; helping others is usually associated with reduced fitness. Thus the "survival of the fittest" suggests that altruistic behaviors are less likely to evolve in one-shot game scenarios, since altruists are worse off than defectors, who accrue benefits by exploiting cooperators.  This poses an evolutionary conundrum: why do we cooperate? how is cooperation compatible with the "survival of the fittest"?~\cite{axelrod1984evolution,fehr2004social}. Some studies have provided theoretical and/or experimental evidence that altruistic punishers--i.e., those who both cooperate and punish defectors--are crucial for the emergence and maintenance of cooperation in one-shot games~\cite{west2007social,henrich2006costly,fehr2002altruistic}. In the one-shot social dilemma game, individuals are required to make their choices simultaneously: they can either benefit themselves by retaining their contributions, or benefit collectives by fully contributing. Within the context of one-shot game scenarios, experimental results have found that~\cite{fehr2002altruistic}: (i) in the absence of any reciprocity mechanism, the average contribution starts at an intermediate level but declines gradually as the game proceeds; and (ii) if costly punishment is available for individuals, the average contributions increase gradually and can be stabilized at a certain level. Altruistic punishers have been identified in large-scale economic and field studies, and their cooperation-enhancing effects have been validated both experimentally and theoretically ~\cite{henrich2006costly,balafoutas2012norm,balafoutas2014direct,guala2012reciprocity,baumard2010has}. However, the abundance of altruistic punishment also poses an evolutionary puzzle: why do we punish and how can such an altruistic behavior evolve? 

The biggest impediments to the evolutionary stability of altruistic punishment are second-order free riders and antisocial punishers.~\cite{herrmann2008antisocial,panchanathan2004indirect,fowler2005second}. Second-order free riders cooperate but do not punish defectors~\cite{fehr2004don,milinski2008punisher}, while antisocial punishers do not cooperate themselves but punish other cooperators. The antisocial punishment strategy has also been observed experimentally in different human cultures~\cite{herrmann2008antisocial}. According to the principle of the "survival of the fittest," second-order free riders are better off than altruistic punishers, so the altruistic punishers should be penalized during evolution, and thus selection should favor defectors to dominate in the whole population. The motivations for antisocial punishment are usually retaliatory punishment or simply targeted to cooperators. Solutions for these two problems can be branched into two research lines. One method involves spatial structures in which each agent can only interact with its direct neighbors~\cite{helbing2010evolutionary,helbing2010punish,nakamaru2005evolution,nakamaru2006coevolution,sekiguchi2009effect,perc2012self,brandt2003punishment,szolnoki2017second}. The spatial structure approach makes it possible to avoid direct competition between prosocial punishers and second-order free riders, and prosocial punishers can therefore avoid being wiped out by second-order free riders~\cite{brandt2003punishment}. However, if antisocial punishment is possible, the problems of second-order free riding and antisocial punishment can cancel each other out in an unlikely and counterintuitive evolutionary outcome, which restores the effectiveness of prosocial punishment~\cite{szolnoki2017second}. The other line of research is based on social mechanisms, such as reputation~\cite{panchanathan2004indirect,santos2011evolution,rockenbach2006efficient,ohtsuki2009indirect,hilbe2012emergence}, coordination or prior commitment~\cite{garcia2019evolution,han2016emergence,boyd2010coordinated}, voluntary participation~\cite{brandt2006punishing,mathew2009does,hauert2007via,fowler2005altruistic,rand2011evolution}, group selection~\cite{boyd2003evolution,saaksvuori2011costly,gavrilets2014solution}, and others~\cite{henrich2001people,amor2011effects}. Although these social mechanisms appear to be materially different, they share a common basis in positive assortativity, which is generally interpreted as the mutual recognition of cooperators. Existing studies of the conundrum of altruistic punishment require the mutual recognition of cooperators, however, the gradual expansion of social dilemma games with punishment brings a degree of complexity into the model.

As artificial intelligence (AI) continues to improve, humans have developed use cases in which AI can benefit the society. This is possible since AI has already been shown to surpass human capabilities in many endeavors. Recently, some scholars have used AI to study the problem of fairness or to investigate the problem of cooperation by studying social dilemma games in hybrid human/AI populations~\cite{santos2019evolution,ishowo2019behavioural}. However, to the best of our knowledge, the influence of AI on humans has not been studied in the context of the conundrum of punishment, especially in the copresence of the two biggest impediments of prosocial punishment mentioned above. Within a framework of a one-shot game, here we explore the conundrum of prosocial punishment by asking several key questions, namely: can AI help humans to establish stable prosocial punishment? How does the spatial configuration of the bots affect the evolutionary outcome? How many bots are needed to establish the dominance of prosocial punishment in a model human society?

To address these questions, we propose a novel human-machine game model based on evolutionary game theory. We chose the two-stage prisoner's dilemma game as our basic model, in which players must decide whether to cooperate with an opponent in the first stage and whether to punish the same opponent in the second stage. This game was modeled in a one-shot setting. In the absence of any reciprocity mechanism, the one-shot version of the two-stage prisoner's dilemma game has been argued to be among the most challenging environments for the emergence of prosocial punishment; this makes it deal for the study of prosocial punishment~\cite{szolnoki2017second,han2016emergence}. We extended the two-stage prisoner's dilemma game by introducing simple bots to a population of human players. Human participants simply play with an opponent and update their actions via social learning without any prior knowledge about the opponent. Since the situation implemented here was a full anonymous situation, in which all prior information is fully excluded, bots can therefore be designed to always perform a specific action, or to always perform an action with a fixed probability. We begin our analysis in an infinite and well-mixed population, in which the probability of normal players playing with bots is the same.
In contrary to well-mixed settings, a networked population enables local interactions among individuals, with players limited to interacting only with their nearest neighbors. We thereafter investigated the evolution of prosocial punishment in two representative networks: a regular lattice, which is known to be simple but possesses the fundamental characteristics of social networks; and a scale-free network that have a heavy-tailed degree distribution. Using these models, we find simple bots are essential for establishing prosocial punishment in both well-mixed and networked populations. In particular, we found that if bots were placed on hub nodes in scale-free networks or if humans possess a psychological bias toward a "copy the majority" rule on a regular lattice, then the dominance of prosocial punishment can be easily to achieved.   

\begin{figure*}[htb]
    \centering
    \includegraphics[width=0.81\linewidth]{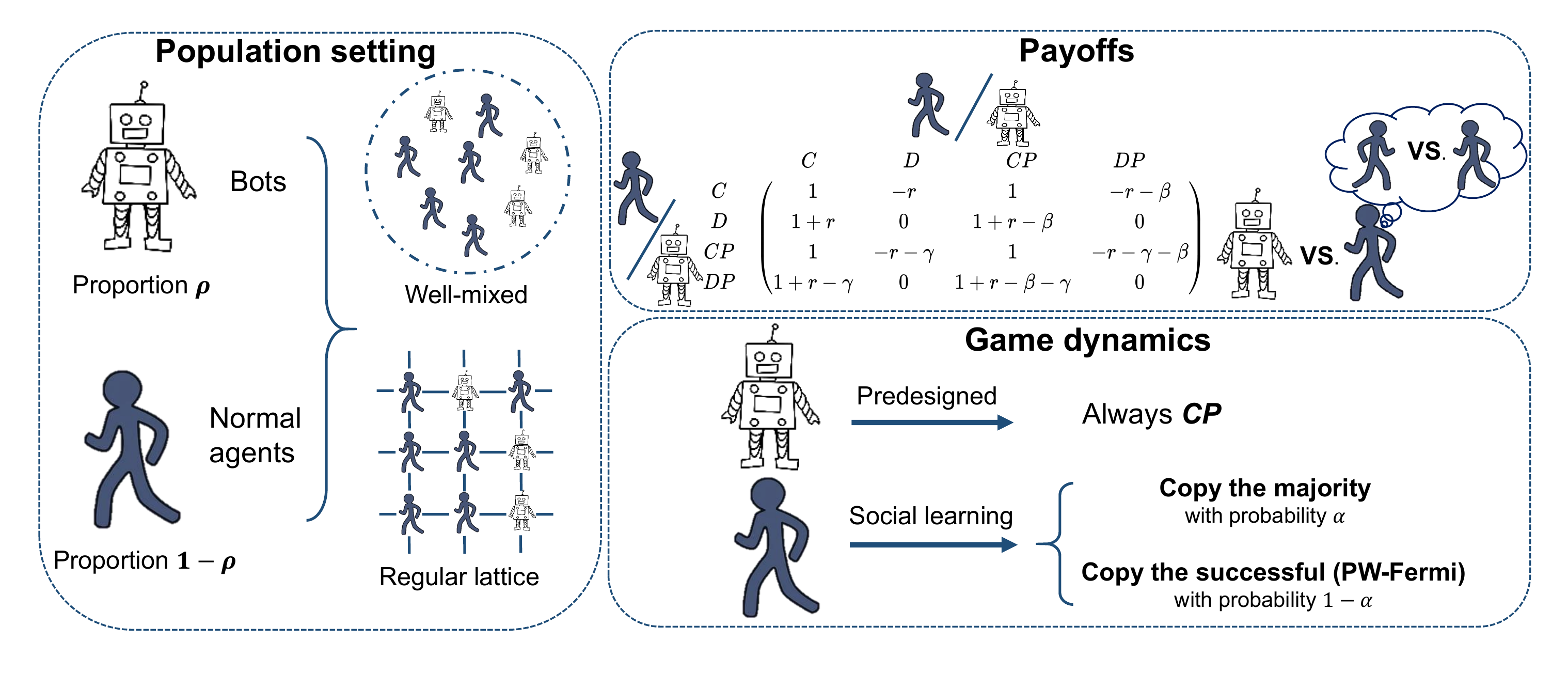}
      \caption{\textbf{Schematic illustration of the human-machine game.} The whole population (either a well-mixed population or networked populations) contains two types of players: bots (proportion $\rho$) and normal players (proportion $1-\rho$). Each player (both normal player and bot) obtains a payoff by interacting with opponent according to the payoff matrix defined in Table. \ref{t01}. Bots are designed to always choose action $CP$ and do not change their actions via social learning. At the same time, we assume that normal players update their actions through social learning--i.e., that normal players update their actions by copying the majority action with probability $\alpha$ and copying the successful (pair-wise Fermi rule) action with probability $1-\alpha$. }
    \label{fig_m}
\end{figure*}

\begin{table}
\caption{\label{t01} Payoff matrix for a two-stage donor-recipient game containing four competing actors: traditional cooperators ($C$) and defectors ($D$), prosocial punishers ($CP$) who cooperate and punish defectors, as well as antisocial punishers ($DP$) who defect and punish cooperators.}
\begin{ruledtabular}
\begin{tabular}{ccccc}
~   & $C$ & $D$ & $CP$ & $DP$     \\
\hline
$C$ & $1$ & $-r$ & $1$  & $-r-\beta$\\
$D$ & $1+r$ & $0$ & $1+r-\beta$ & $0$\\
$CP$ & $1$  & $-r-\gamma$ & 1 & $-r-\gamma-\beta$ \\
$DP$ & $1+r-\gamma$ & $0$ & $1+r-\beta-\gamma$ & 0\\
\end{tabular}
\end{ruledtabular}
\end{table}

\section{Methods}
Our method requires four elementary components: (i) a payoff matrix (ii) population settings, (iii) game dynamics, and (iv) simulation settings and robustness checks. We briefly describe each of elements as follows:

\paragraph{Payoff matrix.} We employed a two-stage donor and recipient game as a basic model to capture the essential social dilemma. In the first stage, all players must simultaneously choose between cooperation ($C$) and  defection ($D$). Cooperation means transferring a benefit $b$ to its opponent at an own cost $c$ ($b>c$), while defection implies inaction, and no benefit is transferred or cost is incurred. Mutual cooperation yields a reward $R=b-c$ while mutual defection yields a punishment $P=0$. If one player cooperates and the other defects, the former receives the sucker's payoff $S=-c$ while the latter receives the temptation to defect $T=b$. After rescaling (by presuming $r=c/(b-c)$) and substituting, the payoff matrix yields $T=1+r$, $R=1$, $P=1$, and $S=-r$, which is a special (diagonal) case of the payoff structure of the prisoner's dilemma game. To quantify the extent of the dilemma, we used the concept of universal dilemma strength for the model~\cite{wang2015universal}. Ref. \cite{tanimoto2021sociophysics} confirms that the dilemma strength is equal to: ${D_{g}}'={D_{r}}'= r$. This specific prisoner's dilemma game possesses aspects of both the chicken-type dilemma (which originates from greed) and the  stag-hunt-type dilemma (which originates from fear).

In the second stage, players must decide whether to punish their opponent at a personal cost $\gamma$. The opponent experiences this punishment as the imposition of a fine $\beta$. To avoid the potential influence of immoral punishers (i.e., defectors that punish defectors), which has been argued to be a beneficial action for the emergence of prosocial punishment~\cite{helbing2010punish}, we intentionally leave out immoral punishment and the strange punishment (cooperators that punish cooperators, which is evolutionarily unstable) actions.  We are therefore able to focus solely on the effectiveness of bot actions on humans. The above process yields four elementary actions, which are:
\begin{itemize}[topsep=0pt,parsep=0pt,itemsep=0pt]
\item $C$, cooperate but do not punish. Users of this action are also known as "second-order free riders", since they free ride on punishment by saving the costs of punishing defectors and thus have a higher fitness than prosocial punishers.
\item $D$, defect and do not punish. Users of this action are also known as "first order free riders". 
\item $CP$, cooperate and punish defectors. Users of this action are known as "prosocial punishers". These are cooperators who punish defectors and therefore need to bear the additional cost of punishment. Consequently, they have lower fitness than second-order free riders.  
\item $DP$, defect but punish cooperators. Users of this action are known as "antisocial punishers." Antisocial punishment may arise as revenge after being punished or may simply target cooperators~\cite{herrmann2008antisocial,gachter2011limits}.
\end{itemize}

The actions described above are summarized in Table. \ref{t01}. To investigate how simple bots can affect the evolution of prosocial punishment, we focus on situations when the effectiveness of punishment is weak by fixing $\gamma=0.1$ and $\beta=0.3$. To ensure the basic social dilemma, the dilemma strength $r$ is limited in the range of (0, 1].    

\paragraph{Population settings.} We consider generally two types of populations: a well-mixed population and two networked populations. The former ensures that global interactions are possible such that all individuals can interact with any others. In contrast, the latter enables only local interactions such that all individuals can only interact with their own local neighbors. To capture the intrinsic proprieties of human behavior, we used two representative networks to model these behaviors. For a homogeneous network, the basic network structure used in this paper was a two-dimensional regular lattice with periodic conditions. We also assumed that the degree of the regular lattice was four, in other words, that each node has four nearest neighbors: up, down, left, and right. To create heterogeneous networks, we used the Barab\'{a}si-Albert algorithm to generate scale-free networks for the purpose of simulation~\cite{albert2002statistical}.  

For both well-mixed and networked populations, we assumed the whole population was initially randomly divided into two types. The first type included normal players, and the other type included bots. The normal players could choose between actions $C$, $D$, $CP$, and $DP$ with equal probability, but the bots were designed to always choose action $CP$. 

\paragraph{Game dynamics.} For the infinite and well-mixed population, we employed mean-field theory to obtain the results of the human-machine game. Let $x,y,z$, and $w$ denote the fractions of cooperators ($C$), defectors ($D$), prosocial punishers ($CP$), and antisocial punishers ($DP$) in the population, such that $x+y+z+w=1$. We assume that the fraction of bots is $\rho$, the total population density is $1+\rho$. The expected payoffs of each actor are as follows:
\begin{equation}
\begin{array}{l}
\Pi_{C}  = \frac{x-yr+z+w(-r-\beta)+\rho}{1+\rho} \\
\Pi_D  = \frac{(1+r)x+(1+r-\beta)(z+\rho)}{1+\rho} \\
\Pi_{CP}  = \frac{x+(-r-\gamma)y+z+w(-r-\gamma-\beta)+\rho}{1+\rho}\\
\Pi_{DP}  = \frac{x(1+r-\gamma)+(z+\rho)(1+r-\gamma-\beta)}{1+\rho}
\end{array}.
\label{eq01}
\end{equation}

Here, we adopt the pairwise comparison rule, which notes that the imitation probability depends on the payoff difference between two randomly selected players. If the randomly selected players have chosen the same action, then nothing happens. Otherwise, the probability that action $i$ replaces action $j$ is determined by the pair-wise Fermi rule (PW-Fermi):
\begin{equation}
W_{j\rightarrow{}i}=\frac{1}{1+\exp\left(\left(\Pi_i-\Pi_j\right)/\kappa\right)}.
\label{eq02}
\end{equation}
Where $i\not=j \in \{C,D,CP,DP\}$, and $\kappa^{-1}>0$ is the imitation strength and it measures how strongly the players are basing their decisions on payoff comparisons~\cite{szabo1998evolutionary,sigmund2010social}. Note that as $\kappa^{-1}\rightarrow+\infty$, players always imitate the most successful action since $W_{i\leftarrow{}j}=1$ if $\Pi_i<\Pi_j$, and $W_{i\leftarrow{}j}=0$ if $\Pi_i>\Pi_j$. While as $\kappa^{-1}\rightarrow0$ (or when $\Pi_i=\Pi_j$), players update their action at random since since $W_{i\leftarrow{}j}=0.5$ in this case. We therefore denote small values of $\kappa^{-1}$ (i.e., $\kappa\rightarrow0$) as the regime of weak imitation, while large values of $\kappa^{-1}$ (i.e., $\kappa\rightarrow+\infty$) were denoted as the regime of strong imitation. For simplicity, we fix the value of $\kappa$ to $\kappa=0.1$ throughout the body text, in addition, we will also investigate how imitation strength affects evolutionary outcomes in the Supplementary Information.

The evolutionary dynamics for the well-mixed and infinite population under the imitation rule can be represented by:
\begin{equation}
\begin{array}{l}
\dot{x} = \frac{2}{1+\rho}(xyP_{D \to C} + xzP_{CP \to C}+xwP_{DP \to C}
-xyP_{C \to D}\\
-x(z+\rho)P_{C \to CP}-xwP_{C \to DP})\\
\dot{y} = \frac{2}{1+\rho}(xyP_{C \to D} + yzP_{CP \to D}+ ywP_{DP \to D}
-xyP_{D \to C}\\
-y(z+\rho)P_{D \to CP}-ywP_{D \to DP})\\
\dot{z} = \frac{2}{1+\rho}(x(z+\rho)P_{C \to CP} +y(z+\rho)P_{D \to CP}+ w(z+\rho)\\
P_{DP \to CP}
-xzP_{CP \to C}-yzP_{CP \to D}-wzP_{CP \to DP})\\
\dot{w} = \frac{2}{1+\rho}(xwP_{C \to DP} +ywP_{D \to DP}+ wzP_{CP \to DP}\\
-xwP_{DP \to C}-ywP_{DP \to D}-w(z+\rho)P_{DP \to CP})\\
\end{array}.
\label{eq03}
\end{equation}
Due to the existence of exponential functions and the high dimensionality of the action space, the stability analysis for Eq. \ref{eq03} is difficult. We therefore use the Runge-Kutta method to generate numerical solutions.

In networked populations, the probability that each node was designated as a bot is $\rho$, and the corresponding probability that each node was designated as a normal player is $1-\rho$. All bot were designed to be prosocial punishers ($CP$), that is, they always cooperates with opponents and punish opponents if they defect. For the population of normal players, each player was designated as a traditional cooperator ($C$), traditional defector ($D$), prosocial punisher ($CP$), or antisocial punisher ($DP$) with equal probability. Normal players and bots residing on networks obtained their payoffs by interacting with all direct neighbors. For each round of the game, a randomly selected normal player $x$ updated its action by imitating the action of a direct neighbor (including bots and normal players) $y$ selected at random using a probability determined by the pair-wise Fermi function (PW-Fermi), which is defined in Eq. \ref{eq02}.

The pair-wise Fermi imitation rule assumes that, during the social learning process, individuals only care about how they can maximize their own personal interest. Consequently, individuals update their action by copying the most successful action (i.e., the action with the highest payoff). However, both theoretical and experimental studies suggest that humans may copy not only the most successful strategy but potentially also the most common strategy~\cite{henrich1998evolution,takahasi1999theoretical}. The importance of conformist transmission in human social learning processes has also been addressed in the literature~\cite{boyd1985culture,henrich1998evolution,henrich2001people}. In our model, we incorporated conformist transmission into the social learning process by assuming that a normal player updates its action by imitating the most common action with probability $\alpha$, and imitates the most successful action with probability $1-\alpha$. The parameter $\alpha$ denotes the strength of conformist transmission, and we assume that $\alpha$ varies from 0 to 1, because otherwise the fittest action cannot spread and the results become meaningless. When $\alpha=0$, the situation reflects only the PW-Fermi situation rule, and the larger the value of $\alpha$ the stronger the strength of conformist transmission. The above procedures are summarized in Figure. \ref{fig_m}.

\paragraph{Simulation settings} We use Monte Carlo simulations to obtain simulation results. A full Monte Carlo simulation was to repeat the above game dynamics for $L\times L$ times ($L^2$ is employed network size.), such that each player can update its action once on average. Typically, we fixed the value of $L$ at 100 throughout the study. We generally ran all simulations for 100,000 steps and obtained the fraction of each actor at an equilibrium state by averaging the last 5,000 steps. In addition, to eliminate the influence of initial conditions, data were averaged over 20 independent runs. 

\paragraph{Robustness checks.}
We checked the robustness of these results by varying the action updating rules, network size, the variant of bots, imitation strength, and the mutation rate. We also varied the lattice size from $L \times L=50^2$ to $L \times L=400^2$, the imitation strength from $\kappa=10^{-4}$ to $\kappa=10^{2}$, and the mutation rate from $\mu=10^{-10}$ to $\mu=10^{0}$. 

\begin{figure*}
    \centering
    \includegraphics[width=0.81\linewidth]{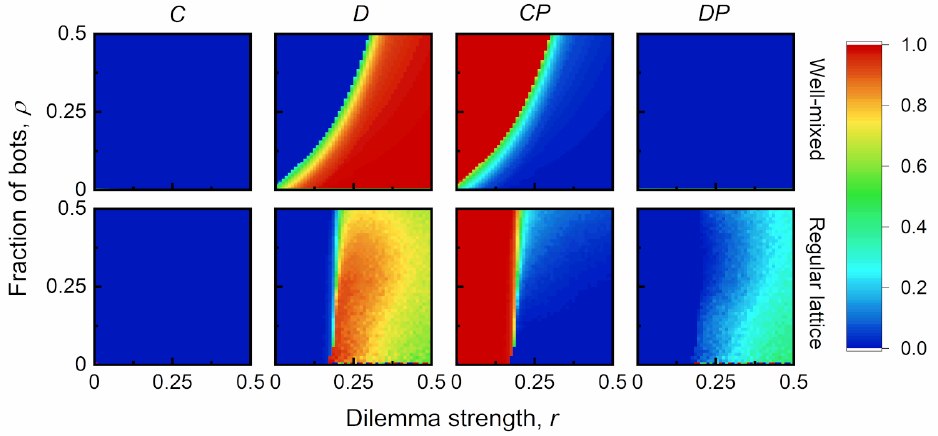}
      \caption{\textbf{Simple bots open up a new pathway to prosocial punishment among humans.} Shown are the fractions of each actor among normal players at steady state as a function of dilemma strength $r$ and bot density $\rho$ in a well-mixed population (top row) and a regular lattice (bottom row). From left to right, the color heatmap indicates the fraction of players using the cooperation, defection, prosocial punishment, and antisocial punishment actions, respectively. We used mean-field theory and Monte Carlo simulation to generate the stationary fractions of each actor for a well-mixed population and a regular lattice, respectively. Results were obtained solely under the PW-Fermi imitation rule, and $\alpha$ was fixed as 0.}
    \label{fig1}
\end{figure*}

We used the myopic best response rule~\cite{matsui1992best,blume1993statistical,roca2009promotion} to check the robustness of our results. In contrast to the PW-Fermi rule, the myopic best response is an innovation rule that allows extinct actions to appear again in the population, while PW-Fermi imitation rule cannot do that. In addition, the myopic best response rule only requires individuals to update their actions based on the previous actions of their neighbors, and hence makes only modest requirements on the cognitive capabilities of individuals. More precisely, a randomly selected focal player $x$ with action $s_x$ updates its action with the following probability:
\begin{equation}
W_{s_x\leftarrow{}s_{x^{'}}}=\frac{1}{1+\exp\left(\left(\Pi_{s_x}-\Pi_{s_x^{'}}\right)/\kappa\right)},
\label{eq05}
\end{equation}
Here, $\Pi_{s_x}$ indicates the payoff of player $x$ with action $s_x$ by playing with all its direct neighbors. $\Pi_{s_x^{'}}$ is the payoff of the same player $x$ if they adopt another random action $s_x^{'}$ within the same neighborhood. Action $s_x^{'}$ is different from action $s_x$, and it is drawn randomly from the remaining three actions. 

The variants of the bots were designed as follows: bots with action $C$ ($B\_C$), bots with action $D$ ($B\_D$), and bots with action $DP$ ($B\_DP$). In addition, we also considered a scenario that included diverse bots action ($B\_ALL$), in which four different bots ($B\_C$, $B\_D$, $B\_CP$, and $B\_DP$) were introduced into the population with equal probability $\frac{\rho}{4}$. We then checked the effect of bots variants on evolutionary outcomes and investigated which bot variant was most beneficial for the dominance of $CP$.

\section{Results}
Here, we first briefly review the results regarding prosocial punishment in the presence of antisocial punishment in a well-mixed population versus a networked population. We then turn our analysis to hybrid populations containing both normal players and bots. 

In one-shot game, if antisocial punishment is allowed in the second-stage of the game, the survival of the cooperative actors (either $C$ or $CP$) is only possible under the most favorable conditions (i.e., weak dilemma strength) in both well-mixed and networked populations~\cite{szolnoki2017second,rand2010anti,han2016emergence}. In a well-mixed population, evolutionary outcomes largely depend on byproduct mutualism, and cooperative actors can dominate the population only when the dilemma strength is weak. In contrast, when the dilemma is relatively strong, the effectiveness of prosocial punishment can be destroyed by second-order free riders and antisocial punishment~\cite{rand2010anti,han2016emergence}. In networked populations, the survival of cooperative actors is only possible when the cost-to-fine ratio is favorable for prosocial punishment~\cite{szolnoki2017second}. As we will show later, this situation significantly changes when we consider hybrid populations containing humans and bots.
\begin{figure*}
    \centering
    \includegraphics[width=0.78\linewidth]{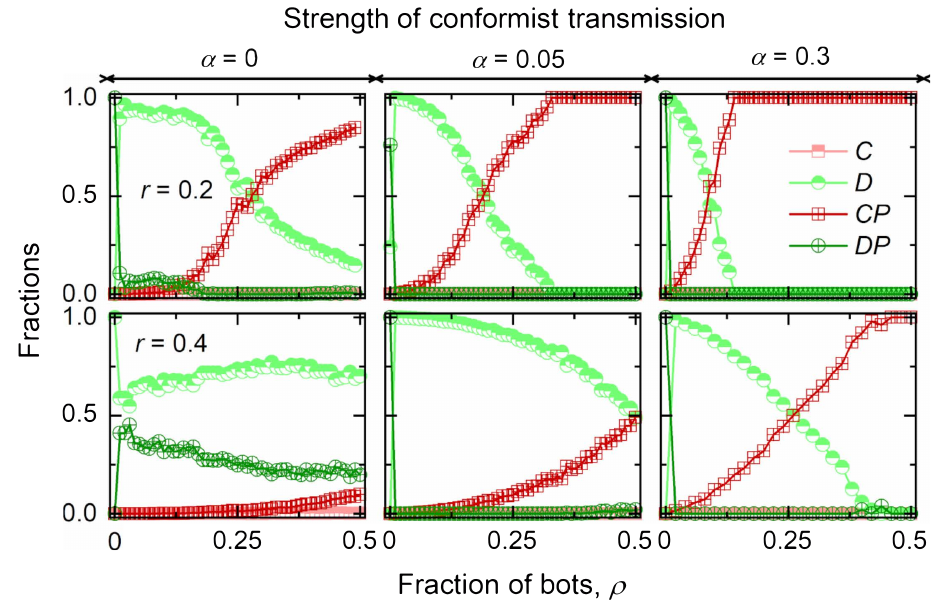}
      \caption{\textbf{If normal players have a psychological bias toward a "copy the majority" rule, weak conformist transmission improves prosocial punishment among normal players, and strong conformist transmission makes this improvement much more significant}. Shown are the abundances of each actor as a function of bot density of bots under three situations: a scenario including only a "copy the successful" rule (left), a weak conformist transmission scenario (middle), and a strong conformist transmission scenario (right). We fixed the strength of the conformist transmission $\alpha$ at 0, 0.05, and 0.3 from left to right, respectively. Dilemma strength was fixed as $r=0.2$ and $r=0.4$ for the top and bottom rows, respectively. }
    \label{fig2}
\end{figure*}

Figure.\ref{fig1} presents the phase diagrams of the densities of each actor as a function of dilemma strength $r$ and the fraction of bots $\rho$. Results are shown for 
two situations: a well-mixed population (top row) and a regular lattice (bottom). In the absence of simple bots (i.e., $\rho=0$), cooperative actors can be maintained but never dominate in an infinite and well-mixed population. The maintenance of cooperative actors largely depends on the initial fractions of the prosocial punishment: only when the initial fraction of prosocial punishers exceeds a certain level can the population be dominated by prosocial punishment (figure.~\ref{figs1}). In a regular lattice, however, if $r\lesssim 0.17$, prosocial punishment can dominate the whole population with the support of traditional network reciprocity. This is in sharp contrast with results for an infinite and well-mixed population (figure.~\ref{figs2}). If simple bots, predesigned to act using action $CP$, were added to either an infinite and well-mixed population or a regular lattice, the fraction of prosocial punishment is significantly improved, and prosocial punishment can be maintained in a more sizable regions of dilemma strength $r$. In addition, the minimum fraction of bots required for the dominance of $CP$ also increases with increasing dilemma strength. Interestingly, this improvement effect in an infinite and well-mixed population is more significant than in a regular lattice. For example, when bot density reaches its maximum value (i.e., $\rho=0.5$), prosocial punishment actually dominants the whole population if the dilemma strength $r$ is less than 0.3 in an infinite and well-mixed population. While a regular lattice scenario shrinks the dominance region of prosocial punishment to $r \lesssim 0.2$ (figure.~\ref{figs2}). It is also worth to note that the the promotion effect of bots to prosocial punishment is only limited to the weak imitation scenario ($\kappa^{-1} \lesssim 10$), while strong imitation scenario diminishes this promotion effect, and leads the population dominated by defective actors (figure.\ref{figs2.1} in SI). 

We also found that, bots promote prosocial punishment among normal players due to a "sticky effect", in which prosocial punishers at nodes near to the bots receive sufficient help to survive even when threatened with second-order free riders and antisocial punishment (SI, left column in figure.\ref{fig3}). However, established prosocial punishers are less effective in convincing their neighbors to adopt the same action (SI, left column in figure.\ref{figs3}). If bots are assigned in regular lattice at random, local interactions generate heterogeneous interaction probabilities between normal players and bots. The probabilities in turn cannot maximize the adhesion effect of bots. Moreover, to some extent they hinder the prosperity of prosocial punishment compared to well-mixed populations characterized by homogeneous interaction probabilities. If the maximum interaction probability between humans and bots is 
guaranteed in a regular lattice, then the prosocial punishment level can be relatively high; i.e., well beyond the levels found in of random distributions (SI, bottom right column in figure.\ref{figs3} and left column in figure.\ref{fig3}). Moreover, large interaction probabilities between bots and humans generally lead to an optimal level of prosocial punishment (figure.~\ref{figs3} in SI). However, the spatial configurations of bots matter. Keeping the density of bots and the interaction probability between bots and humans remains unchanged, different spatial configurations of bots can produce quite different prosocial punishment levels (figure.\ref{fig3} in SI).

\begin{figure*}
    \centering
    \includegraphics[width=0.78\linewidth]{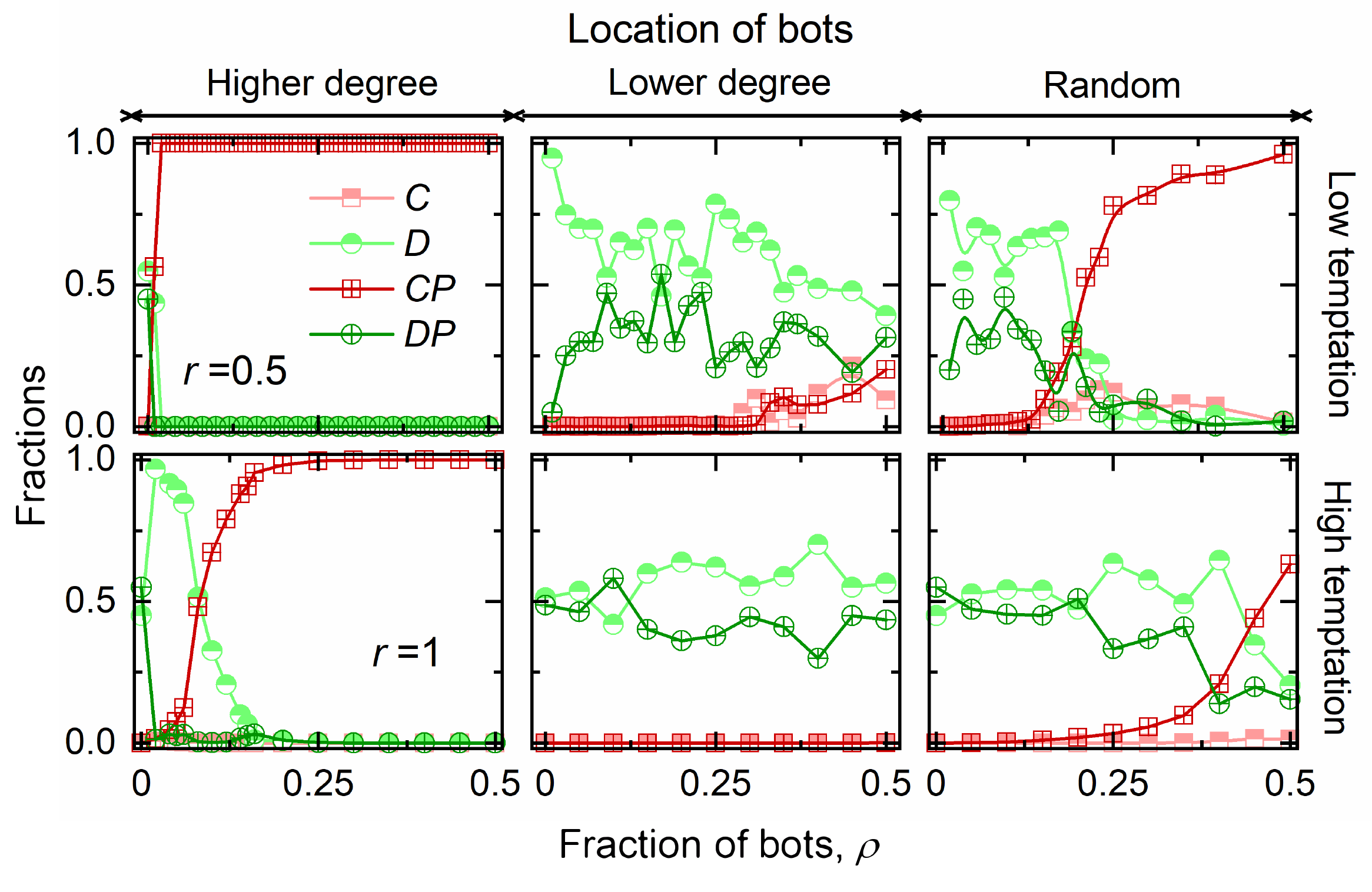}
      \caption{\textbf{Bots located at higher-degree nodes cause a significant improvement in the prosocial punishment level, and the complete dominance of prosocial punishment requires only a small fraction of bots.} Shown are the fractions of each actor independent of the density of bots under three different situations: (i) when bots were preferentially allocated to higher-degree nodes (left), (ii) when bots were preferentially allocated to lower-degree nodes (middle), and (iii) when bots were allocated to scale-free networks at random (right). We used dilemma strength $r=0.5$ and $r=1$ for the top and bottom rows, respectively. Results were obtained under the pair-wise Fermi rule alone.}
    \label{fig4}
\end{figure*}

To some extent, the effect of bots to promote prosocial punishment is related to the upstream indirect reciprocity or social contagion~\cite{nowak2007upstream,boyd1989evolution,christakis2013social,fowler2010cooperative}. Both of these theories refer to the situation in which a person has just received help and has a large probability to pass this help on to a third person~\cite{nowak2007upstream}. In particular, the social contagion view of cooperation suggests that cooperation behaviors cascade in human social networks, with cooperative behavior being able to spread up to three degrees of separation from a local agent~\cite{fowler2010cooperative}. On a collective level, bots that predesigned to choose action $CP$ directly increase the probability that a normal player receives help, but this cannot guarantee that there is a greater probability that the person who was just helped will pass on this help to a third person. The improvement of the probability in the latter case requires weak imitation (or weak selection)~\cite{traulsen2010human,kirchkamp2007naive}, whereby individuals do not care too much about material gains when updating their actions (see figure.\ref{figs2.1} in SI). However, if we relax the assumption that individuals must follow the "copy the successful" rule in the simulation, and allow individuals to also follow the "copy the majority" rule, then the prevalence of prosocial punishment among humans can be more easily achieved compared that if the "copy the successful" rule is the only option.

Figure.\ref{fig2} shows the fractions of each actor as a function of bot density. Shown are both weak (top row) and strong (bottom row) dilemma strength scenarios. Each is shown under three different situations: one only containing the "copy the successful" rule (left column in figure.\ref{fig2}), a weak conformist transmission scenario (middle column in figure.\ref{fig2}), and a strong conformist transmission scenario (right column in figure.\ref{fig2}). Under the scenario containing only the "copy the successful" rule (left column in figure.\ref{fig2}), if the dilemma strength is weak (i.e. $r=0.2$), increasing the density of bots significantly improves the prosocial punishment level, but $CP$ cannot fully dominate throughout the population even for largest bot density (top left panel in figure.\ref{fig2}). The high dilemma strength makes the effort to increase $CP$ futile, and the maximum level of $CP$ is less than 10\% when the density of bots reaches its maximum value (bottom left in figure.\ref{fig2}). However, if players have a learning bias toward a "copy the majority" rule in addition to a "copy the successful" rule, increasing the density of bots always leads to significant improvement in the prosocial punishment level. In particular, for the weak conformist transmission scenario (i.e. in which $\alpha=0.05$), fully dominance of $CP$ requires just 30\% of bots under weak dilemma strength conditions, and the $CP$ level can also reach 50\% even under high dilemma strength conditions (middle column in figure.\ref{fig2}). The strong conformist transmission scenario makes these improvements much more significant. Here, the dominance of $CP$ requires only 15\% and 40\% of bots for weak and strong dilemma strength conditions, respectively (right column in figure.\ref{fig2}). These observations are robust against different values of the dilemma strength (SI, see figure~\ref{figs11} and figure.\ref{figs12}).

It is known that cooperative clusters are the hall-marks of network reciprocity, through which cooperators can survive defectors invasions. As mentioned above, the introduction of bots causes $CP$ to proliferate among normal players, but the established prosocial punishers rarely convince their neighbors to adopt $CP$ if players care only about material gains. In contrast, if players have a learning bias toward a "copy the majority", especially in a strong conformist transmission scenario, the area of $CP$ adoption around bots can grow to form large, compact clusters that support each other (see: SI, comparisons between the first, third and last columns of figure.\ref{figs4}). Thus the rate of $CP$ adoption is higher than under a scenario where normal players have only a "copy the successful" rule, and finally brings enhanced networked reciprocity to the population. This means that in a homogeneous network, the introduction of bots first causes the spread of prosocial punishment behavior among normal players, but conformist transmission further improves the transmissibility of the establishment of prosocial punishment. Finally, even in the presence of second-order free riders and antisocial punishers, prosocial punishment can dominate the whole population even at a high dilemma strength (i.e. $r \gtrsim 0.17$), while the population will be dominated by either defectors or antisocial punishers when bots are absent. To investigate how network structures affect evolutionary outcomes, we further examine these evolutionary dynamics on scale-free networks. 

In contrast to homogeneous networks, scale-free networks with heavy-tailed degree distributions are characteristic of many socioeconomic systems~\cite{broido2019scale,holme2019rare}. Scale-free networks can capture the intrinsic interactions of humans and have been found to be an optimal network for the maintenance of cooperation due to the outsized effects of hub nodes~\cite{santos2005scale,santos2008social}. We therefore select this network form to represent social networks and to thereby allow us to investigate how heterogeneous degree distributions affect the evolutionary outcomes identified thus far. Consistent with the conclusions reported above, we found that allocating prosocial bots to scale-free networks expands the survival regions of prosocial punishment, and a larger density of bots always leads to better outcomes. The only difference is that--without adhering to conformist transmission--the bots themselves can produce a much more pronounced improvement effect on prosocial punishment in scale-free networks than in regular lattice ( figure.\ref{figs5}).

To further investigate how bot spatial configuration affects the prevalence of prosocial punishment, figure.\ref{fig4} depicts the abundance of each actor independent of bot density under three different situations: (i) a scenario in which bots were preferentially placed on higher-degree nodes, (ii) a scenario in which bots were preferentially placed on lower-degree nodes, and (iii) a scenario in which bots were placed in scale-free networks at random. We observe that when bots were assigned preferentially to lower-degree nodes, bots show only a slight ability to promote prosocial punishment and prosocial punishment therefore cannot dominate the population. Allocating bots to the network randomly leads to a significant improvement in the prevalence of prosocial punishment,
and the dominance of prosocial punishment requires populations that are 25\% and 50\% bots for weak and strong dilemma strengths (top right and bottom right panels of figure.\ref{fig4}), respectively. However, if bots are assigned preferentially to higher degree nodes, there is significant and noticeable improvement in the prevalence of prosocial punishment compared to the other two scenarios. The bot density required for the dominance of prosocial punishment shrinks to 2\% and 10\% for relatively weak and strong dilemma strength (top left and bottom left panels of figure.\ref{fig4}). In short, if the bots are assigned to higher-degree nodes, thereby giving them disproportionately strong influence relative to other players, this power can shape attitudes toward prosocial punishment among humans much more easier. Moreover, the dominance of the prosocial punishment only requires a bot density of roughly 10\%.

\section{Conclusions and discussions }
To conclude, we propose a novel evolutionary framework to address the problems of second-order free ridership and antisocial punishment. In contrast to previous solutions developed for the one-shot game, in which the dominance of prosocial punishment requires either voluntary participation or prior commitment~\cite{han2016emergence,hauert2007via,rand2011evolution}, we extend the theory of social punishment by introducing a known fraction of fixed-action bots to the pool of human players. Within the context of the one-shot game, the bot behavior was designed to always $CP$ (cooperate and punish defectors); that is, the bots always maintained the same $CP$ action and never changed. At the same time, we assume that humans do not know whether their opponents are bots or not and also that they do not know the actions of the bots will choose. During evolution scenarios, humans gain a payoff by interacting with opponents, which can be either another human or a bot, and can change their actions through social learning. 

In the absence of bots, we find that,  prosocial punishment cannot survive by itself if the dilemma strength is relatively high (i.e., $r \gtrsim 0.17$ for both regular lattice and well-mixed populations). However, adding a some fraction of bots into the human population not only causes the proliferation of prosocial punishment behavior among humans but also allows the dominance of prosocial punishment even at high dilemma strength. On the other hand, the effect of the bots on humans is also related to the initial spatial configuration of the bots within the available network structure. In regular lattice, the optimal level of prosocial punishment can be achieved if humans have the maximum probability to interact with bots. If humans have a psychological bias toward a "copy the majority" strategy, then conformist transmission can improve the transmissibility of the establishment of the prosocial punishment. This result in significant improvement in the final prosocial punishment level. Moreover, if these bots are assigned to scale-free networks, the improvement effect of bots becomes much more significant than in well-mixed or lattice-structured populations. In such a case prosocial punishment can even dominate the entire population over the full range of dilemma strengths without adhering to the "copy the majority" rule. In particular, if the bots were allocated preferentially to higher-degree nodes, the dominance of prosocial punishment requires a bot density of roughly 10\% over the entire range of dilemma strength; this is far beyond the capability of the bots in either well-mixed or regular lattice populations.   

Our results are not limited to an "copy the successful" rule, and is robust against social learning rules. If humans do not rely on an "copy the successful" rule or a "copy the majority" rule, and instead rely on a "myopic best response" rule, our conclusions basically hold. However, this rule produces less significant improvement in the prevalence of prosocial punishment (see the third column of figure.\ref{figs6} in SI). Our results are also robust against network size and mutation. Changing the network size produces little difference to the results obtained on a lattice with $100\times100$ nodes. (figure.\ref{figs7}). The current results also generally hold for small mutation rate, since increasing the mutation rate generally destroys the dominance of $CP$ and leads the system to a full defective state (figure. \ref{figs8}). 

From the perspective of bot design, the behavior of the bots can be designed as either $C$, $D$, $CP$, or $DP$. However, we found that only bots that consistently choose $CP$ can produce significant improvement in the level of prosocial punishment in different types of networks and under different updating rules. Introducing bots that choose action $C$, however, promotes cooperation only in well-mixed populations and scale-free networks (see the first columns of figure.\ref{figs5} and figure.\ref{figs9}), but not in a regular lattice-structured population. In regular lattice, a previous study reported that the existence of zealous cooperators has the effect of destroying rather than boosting cooperation~\cite{matsuzawa2016spatial}. Here, we found this argument generally holds for the conundrum of prosocial punishment. Allocating bots that choose action $C$ harms cooperation under the pairwise Fermi rule (see the first column of figure.~\ref{figs10}), while a scenario that uses the "myopic best response" rule leads the bots to produce little effect on cooperation (see the first column in figure.\ref{figs6}). Introducing defective bots (i.e., bots with action choices of either $D$ or $DP$) always destroys cooperation; this finding is robust against different network types and different updating rules (see the second and the fourth columns of figure.\ref{figs5}, figure.\ref{figs6}, and figure.\ref{figs9}-\ref{figs12}). Moreover, a diversity of bot behavior (tested by adding all four kinds of bot into the population with equal probability) also destroys the cooperation-promotion effect brought by cooperative bots (i.e., those with either action $C$ or $CP$). 

Our results also demonstrate the importance of cultural transmission and the identification of vital nodes~\cite{henrich2001people,lu2016vital}, since we showed that conformist transmission and hub nodes are two critical determinants of the dominance of prosocial punishment. In addition, from the perspective of developing realistic applications of these findings, it is notable that powerful governments and organizations often use zombie accounts within online spaces to influence people's opinions, behaviors, or beliefs. Such zombie accounts can be regarded as a kind of bot through which powerful institutions can build a harmonious society. However, if zombie accounts are maliciously exploited by bad actors, then these zombie accounts may reinforce negative behaviors. Our results relate to concerns about use of zombie accounts, since we show that the level of prosocial behavior is sensitive to bot design. Although bots designed to choose action $CP$ can establish prosocial punishment among human populations, the destruction of prosocial punishment can be much easier, since defective bots always diminish or destroy prosocial punishment among human populations.

The behavioral rules of the bots defined in this paper are same as that of zealots or committed individuals~\cite{centola2018experimental,masuda2012evolution,cardillo2020critical,nakajima2015evolutionary,matsuzawa2016spatial}. 
Although zealots can be found in empirical studies~\cite{coleman1988free}, they are rare among humans in general. Recent empirical studies confirmed that, in studies of social norms, the opinion of the majority can be tipped to that of the minority when roughly 25\% of individuals were highly committed minorities~\cite{centola2018experimental}. In social dilemma games, the required density of zealots to achieve the dominance of cooperation linearly increases with increasing dilemma strength~\cite{masuda2012evolution}. If the maintenance of cooperation or stable social norms requires a large number of zealots, then the cooperation-promotion effect induced by zealots becomes meaningless since it is in conflict with reality. While the concept of bots within the framework of a human-machine game seems more concrete and appropriate, our research targets then become: (i) how to improve human cooperation by adding a small fraction of bots to a human population, and (ii) to determine which kind of bots are most beneficial for human cooperation. The research questions also become more diverse, and then goes beyond the scope of zealotry. For example, within the context of the one-shot game, bots can be simply designed as zealots who always to choose one concrete action. However, if humans can encounter iterated interactions or have some prior information about their opponents, then bot design might become much more complex.    

One critical assumption in our model is that humans do not know whether their opponents are bots or not, i.e., players interact with each other under full anonymous situations. We therefore fully exclude reputation effects and iterated interactions to focus solely on how simple bots can solve the conundrum of prosocial punishment in a one-shot game. Happily, we find that introducing bots with action $CP$ promotes prosocial punishment among human populations, a finding that provides a new explanation for the emergence of prosocial punishment. Recent human behavior experiments have often produced results that were surprising or conflicting when compared to theoretical results. For example, whether peer punishment can promote cooperation~\cite{dreber2008winners,wu2009costly}, whether scale-free topologies can promote cooperation compared to a lattice structure~\cite{gracia2012heterogeneous}, and whether introducing peer punishment to networks can promote cooperation~\cite{li2018punishment}. Therefore, our findings require further experimental confirmation by future studies. In addition, if we relax our assumption and allow humans to know whether their opponents are bots or not, then whether the effect that bots with action $CP$ breed cooperation still holds will largely depend on humans' emotions or attitudes toward bots. In that sense, human behavior experiments will be an efficient tool for determining the emotions of humans toward bots.

\section*{Article information}

\paragraph*{Acknowledgements.} We thank Prof. Dr. Marko Jusup for valuable discussions. We acknowledge support from (i)a JSPS Postdoctoral Fellowship Program for Foreign Researchers (grant no. P21374), and an accompanying Grant-in-Aid for Scientific Research from KAKENHI (grant no. JP 22F31374), and the National Natural Science Foundation of China (grant no.~11931015) to C.\,S. as a co-investigator, (ii) the National Natural Science Foundation of China (grants no.~11931015, ~12271471 and 11671348) to L.\,S., (iii) the National Natural Science Foundation of China (grant no.~U1803263), the Thousand Talents Plan (grant no.~W099102), the Fundamental Research Funds for Central Universities (grant no.~3102017jc03007), and the China Computer Federation--Tencent Open Fund (grant no.~IAGR20170119) to Z.\,W, and (iv) the grant-in-Aid for Scientific Research from JSPS, Japan, KAKENHI (grant No. JP 20H02314) awarded to J.\,T.
\paragraph*{Author contributions.} 
C.\,S. and J.\,T. conceived research. C.\,S. and Z.\, H. performed simulations. All co-authors discussed the results and wrote the manuscript.
\paragraph*{Conflict of interest.} Authors declare no conflict of interest.

\bibliography{biblio}

% \begin{bibunit}
% \putbib
% \end{bibunit}

\clearpage
\onecolumngrid
\setcounter{equation}{0}
\renewcommand\theequation{S\arabic{equation}}
\setcounter{page}{1}
\renewcommand\thepage{S\arabic{page}}

\section*{
Supplementary Information for\\
``Simple bots breed social punishment in humans''}

\noindent\textbf{Remark 1.} Let $x,y,z$, and $w$ denote the fractions of cooperators ($C$), defectors ($D$) prosocial punishers ($CP$), and antisocial punishers ($DP$) in a well-mixed and infinite population such that $x+y+z+w=1$. We assume the fraction of bots is $\rho$, and the bots are designed to always choose $C$ and never change their action. The total population density is then $1+\rho$, and the expected payoffs of the actors are as follows:

\begin{equation}
\begin{array}{l}
\Pi_{C}  = \frac{x-yr+z+w(-r-\beta)+\rho}{1+\rho} \\
\Pi_D  = \frac{(1+r)(x+\rho)+(1+r-\beta)z}{1+\rho} \\
\Pi_{CP}  = \frac{x+(-r-\gamma)y+z+w(-r-\gamma-\beta)+\rho}{1+\rho}\\
\Pi_{DP}  = \frac{(x+\rho)(1+r-\gamma)+z(1+r-\gamma-\beta)}{1+\rho}
\end{array}.
\label{eq01}
\end{equation}

We adopt the pairwise comparison rule, where the imitation probability depends on the payoff difference between two randomly selected players. If the randomly selected players choose the same action, nothing happens. Otherwise, the probability that action $i$ replaces action $j$ is:

\begin{equation}
\begin{array}{l}
P_{j \to i}  = 1/\left[1+e^{(\Pi_{j}-\Pi_{i})/\kappa}\right] \\
\end{array}.
\label{eq02}
\end{equation}

Where $i\not=j \in \{C,D,CP,DP\}$, and $\kappa^{-1}$ is the imitation strength such that $\kappa^{-1}>0$. The evolutionary dynamics of the well-mixed and infinite population under the imitation rule are represented by:
\begin{equation}
\begin{array}{l}
\dot{x} = \frac{2}{1+\rho}((x+\rho)yP_{D \to C} + (x+\rho)zP_{CP \to C}+(x+\rho)wP_{DP \to C}
-xyP_{C \to D}-xzP_{C \to CP}-xwP_{C \to DP})\\
\dot{y} = \frac{2}{1+\rho}(xyP_{C \to D} + yzP_{CP \to D}+ ywP_{DP \to D}
-(x+\rho)yP_{D \to C}-yzP_{D \to CP}-ywP_{D \to DP})\\
\dot{z} = \frac{2}{1+\rho}(xzP_{C \to CP} +yzP_{D \to CP}+ wzP_{DP \to CP}
-(x+\rho)zP_{CP \to C}-yzP_{CP \to D}-wzP_{CP \to DP})\\
\dot{w} = \frac{2}{1+\rho}(xwP_{C \to DP} +ywP_{D \to DP}+ wzP_{CP \to DP}
-(x+\rho)wP_{DP \to C}-ywP_{DP \to D}-wzP_{DP \to CP})\\
\end{array}.
\label{eq04}
\end{equation}

\noindent\textbf{Remark 2.} Let $x,y,z$, and $w$ denote the fractions of cooperators ($C$), defectors ($D$) prosocial punishers ($CP$), and antisocial punishers ($DP$) in a well-mixed and infinite population, such that $x+y+z+w=1$. We assume the fraction of bots is $\rho$, and the bots are designed always to choose $D$ and never change their action. The total population density is then $1+\rho$, and the expected payoffs of each of the actors are as follows:

\begin{equation}
\begin{array}{l}
\Pi_{C}  = \frac{x-(y+\rho)r+z+w(-r-\beta)}{1+\rho} \\
\Pi_D  = \frac{(1+r)x+(1+r-\beta)z}{1+\rho} \\
\Pi_{CP}  = \frac{x+(-r-\gamma)(y+\rho)+z+w(-r-\gamma-\beta)}{1+\rho}\\
\Pi_{DP}  = \frac{x(1+r-\gamma)+z(1+r-\gamma-\beta)}{1+\rho}
\end{array}.
\label{eq01}
\end{equation}

We adopt the pairwise comparison rule, where the imitation probability depends on the payoff difference between two randomly selected players. If the randomly selected players choose the same action, nothing happens. Otherwise, the probability that action $i$ replaces action $j$ is:

\begin{equation}
\begin{array}{l}
P_{j \to i}  = 1/\left[1+e^{(\Pi_{j}-\Pi_{i})/\kappa}\right] \\
\end{array}.
\label{eq02}
\end{equation}

Where $i\not=j \in \{C,D,CP,DP\}$, and $\kappa^{-1}$ is the imitation strength such that $\kappa^{-1}>0$. The evolutionary dynamics of the well-mixed and infinite population under the imitation rule are represented by:
\begin{equation}
\begin{array}{l}
\dot{x} = \frac{2}{1+\rho}(xyP_{D \to C} + xzP_{CP \to C}+xwP_{DP \to C}
-x(y+\rho)P_{C \to D}-xzP_{C \to CP}-xwP_{C \to DP})\\

\dot{y} = \frac{2}{1+\rho}(x(y+\rho)P_{C \to D} + (y+\rho)zP_{CP \to D}+ (y+\rho)wP_{DP \to D}
-xyP_{D \to C}-yzP_{D \to CP}-ywP_{D \to DP})\\

\dot{z} =\frac{2}{1+\rho}(xzP_{C \to CP} +yzP_{D \to CP}+ wzP_{DP \to CP}
-xzP_{CP \to C}-(y+\rho)zP_{CP \to D}-wzP_{CP \to DP})\\
\dot{w} = \frac{2}{1+\rho}(xwP_{C \to DP} +ywP_{D \to DP}+ wzP_{CP \to DP}
-xwP_{DP \to C}-(y+\rho)wP_{DP \to D}-wzP_{DP \to CP})\\
\end{array}.
\label{eq04}
\end{equation}

\noindent\textbf{Remark 3.} Let $x,y,z$, and $w$ denote the fractions of cooperators ($C$), defectors ($D$) prosocial punishers ($CP$), and antisocial punishers ($DP$) in a well-mixed and infinite population, and $x+y+z+w=1$. We assume the fraction of bots is $\rho$, and the bots are designed to always choose $DP$ and never change their action. The total population density is $1+\rho$, and the expected payoffs of the actors are as follows:

\begin{equation}
\begin{array}{l}
\Pi_{C}  = \frac{x-yr+z+(w+\rho)(-r-\beta)}{1+\rho} \\
\Pi_D  = \frac{(1+r)x+(1+r-\beta)z}{1+\rho} \\
\Pi_{CP}  = \frac{x+(-r-\gamma)y+z+(w+\rho)(-r-\gamma-\beta)}{1+\rho}\\
\Pi_{DP}  = \frac{x(1+r-\gamma)+z(1+r-\gamma-\beta)}{1+\rho}
\end{array}.
\label{eq01}
\end{equation}

We adopt the pairwise comparison rule, where the imitation probability depends on the payoff difference between two randomly selected players. If the randomly selected players choose the same action, nothing happens. Otherwise, the probability that action $i$ replaces action $j$ is:

\begin{equation}
\begin{array}{l}
P_{j \to i}  = 1/\left[1+e^{(\Pi_{j}-\Pi_{i})/\kappa}\right] \\
\end{array}.
\label{eq02}
\end{equation}

Where $i\not=j \in \{C,D,CP,DP\}$, and $\kappa^{-1}$ is the imitation strength such that $\kappa^{-1}>0$. The evolutionary dynamics of the well-mixed and infinite population under the imitation rule are represented by:
\begin{equation}
\begin{array}{l}
\dot{x} = \frac{2}{1+\rho}(xyP_{D \to C} + xzP_{CP \to C}+xwP_{DP \to C}
-xyP_{C \to D}-xzP_{C \to CP}-x(w+\rho)P_{C \to DP})\\
\dot{y} = \frac{2}{1+\rho}(xyP_{C \to D} + yzP_{CP \to D}+ ywP_{DP \to D}
-xyP_{D \to C}-yzP_{D \to CP}-y(w+\rho)P_{D \to DP})\\
\dot{z} = \frac{2}{1+\rho}(xzP_{C \to CP} +yzP_{D \to CP}+ wzP_{DP \to CP}
-xzP_{CP \to C}-yzP_{CP \to D}-(w+\rho)zP_{CP \to DP})\\
\dot{w} = \frac{2}{1+\rho}(x(w+\rho)P_{C \to DP} +y(w+\rho)P_{D \to DP}+ (w+\rho)zP_{CP \to DP}
-xwP_{DP \to C}-ywP_{DP \to D}-wzP_{DP \to CP})\\
\end{array}.
\label{eq04}
\end{equation}

\noindent\textbf{Remark 4.} Let $x,y,z,$ and $w$ denote the fractions of cooperators ($C$), defectors ($D$) prosocial punishers ($CP$), and antisocial punishers ($DP$) in a well-mixed and infinite population such that $x+y+z+w=1$. We assume that bots have equal probabilities to be $C$, $D$, $CP$, and $DP$. Thus, the fractions of bots with action $C$, with action $D$, with action $CP$, and with action $DP$ are all $\rho/4$. The total population density is $1+\rho$. Therefore, the expected payoffs of each of the actors are as follows:

\begin{equation}
\begin{array}{l}
\Pi_{C}  = \frac{x+\frac{\rho}{2 }-(y+\frac{\rho}{4})r+z+(w+\frac{\rho}{4})(-r-\beta)}{1+\rho} \\
\Pi_D  = \frac{(1+r)(x+\frac{\rho}{4})+(1+r-\beta)(z+\frac{\rho}{4})}{1+\rho} \\
\Pi_{CP}  = \frac{x+\frac{\rho}{2}+(-r-\gamma)(y+\frac{\rho}{4})+z+(w+\frac{\rho}{4})(-r-\gamma-\beta)}{1+\rho}\\
\Pi_{DP}  = \frac{(x+\frac{\rho}{4})(1+r-\gamma)+(z+\frac{\rho}{4})(1+r-\gamma-\beta)}{1+\rho}
\end{array}.
\label{eq01}
\end{equation}

We adopt the pairwise comparison rule, where the imitation probability depends on the payoff difference between two randomly selected players. If the randomly selected players choose the same action, nothing happens. Otherwise, the probability that action $i$ replaces action $j$ is:

\begin{equation}
\begin{array}{l}
P_{j \to i}  = 1/\left[1+e^{(\Pi_{j}-\Pi_{i})/\kappa}\right] \\
\end{array}.
\label{eq02}
\end{equation}

Where $i\not=j \in \{C,D,CP,DP\}$, and $\kappa^{-1}$ is the imitation strength such that $\kappa^{-1}>0$. The evolutionary dynamics for a well-mixed and infinite population under the imitation rule are represented by:
\begin{equation}
\begin{array}{l}
\begin{split}
    \dot{x} = \frac{2}{1+\rho}((x+\frac{\rho}{4})yP_{D \to C} + (x+\frac{\rho}{4})zP_{CP \to C}+(x+\frac{\rho}{4})wP_{DP \to C}\\
-x(y+\frac{\rho}{4})P_{C \to D}-x(z+\frac{\rho}{4})P_{C \to CP}-x(w+\frac{\rho}{4})P_{C \to DP})\\
\end{split}
\\
\begin{split}
   \dot{y} = \frac{2}{1+\rho}(x(y+\frac{\rho}{4})P_{C \to D} + (y+\frac{\rho}{4})zP_{CP \to D}+ (y+\frac{\rho}{4})wP_{DP \to D}\\
-(x+\frac{\rho}{4})yP_{D \to C}-y(z+\frac{\rho}{4})P_{D \to CP}-y(w+\frac{\rho}{4})P_{D \to DP})\\ 
\end{split}
\\
\begin{split}
    \dot{z} = \frac{2}{1+\rho}(x(z+\frac{\rho}{4})P_{C \to CP} +y(z+\frac{\rho}{4})P_{D \to CP}+ w(z+\frac{\rho}{4})P_{DP \to CP}\\
-(x+\frac{\rho}{4})zP_{CP \to C}-(y+\frac{\rho}{4})zP_{CP \to D}-(w+\frac{\rho}{4})zP_{CP \to DP})\\
\end{split}
\\
\begin{split}
    \dot{w}=\frac{2}{1+\rho}(x(w+\frac{\rho}{4})P_{C \to DP}+y(w+\frac{\rho}{4})P_{D \to DP}+ (w+\frac{\rho}{4})zP_{CP \to DP}\\ 
-(x+\frac{\rho}{4})wP_{DP \to C}-(y+\frac{\rho}{4})wP_{DP \to D}-(z+\frac{\rho}{4})wP_{DP \to CP})\\
\end{split}

\end{array}.
\label{eq04}
\end{equation}

\clearpage
\renewcommand\thefigure{S\arabic{figure}}
\setcounter{figure}{0}  
\section*{Supplementary Figures}
\vfill
\phantom{Invisible text.}

\begin{figure*}[!h]
    \centering
\includegraphics[width=0.58\linewidth]{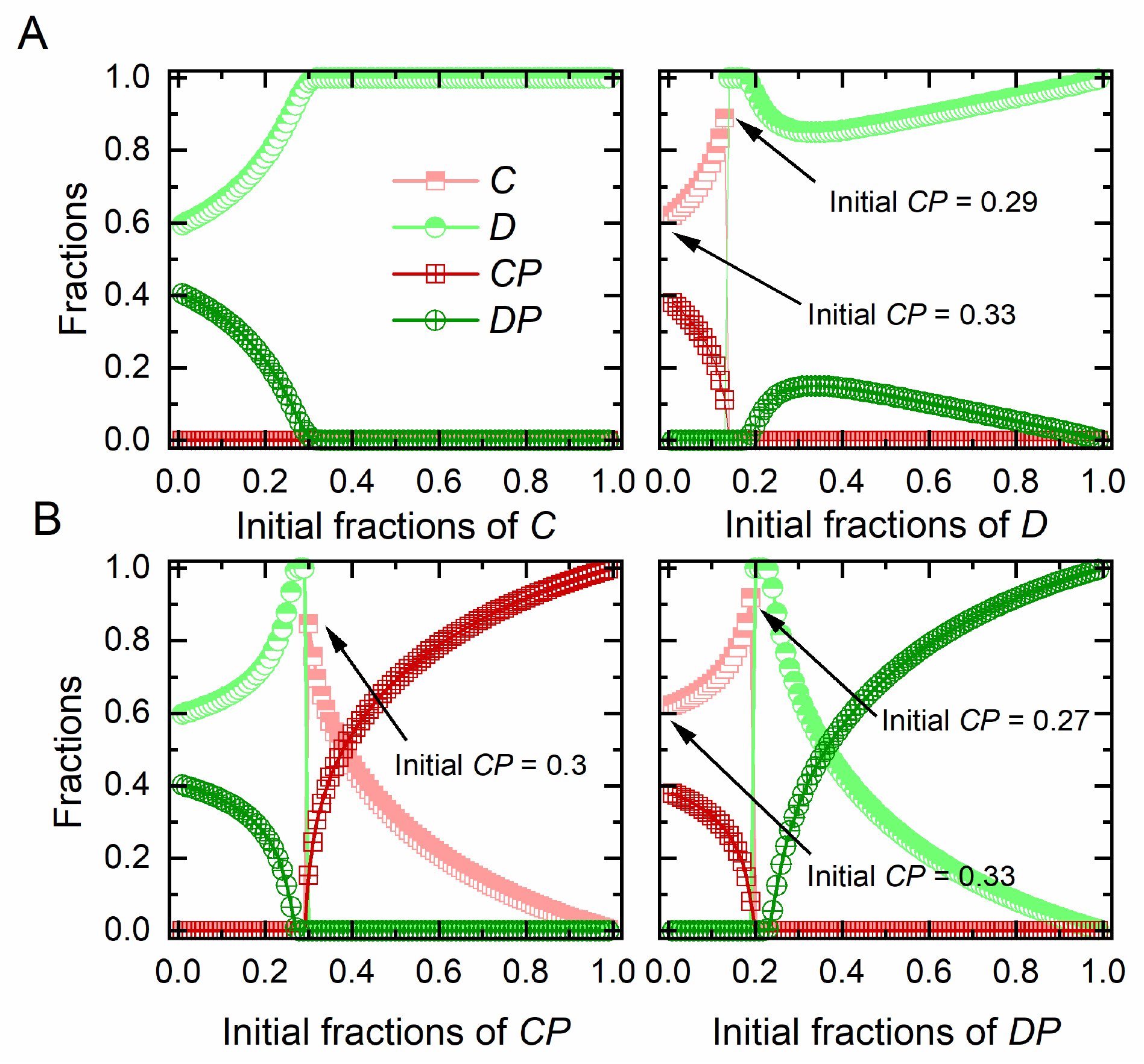}
    \caption{\textbf{In the absence of bots, the fate of altruists largely depends on the initial fractions of prosocial punishers.} Shown are the abundances of each actor as a function of the initial values of cooperators (top left), defectors (top right), prosocial punishers (bottom left), and antisocial punishers (bottom right). We plotted these panels by gradually changing the fixed initial values of one type of actor and equally divided the initial values of the other actor types among the remainder. It is clear that cooperative actors can only emerge if the initial values of prosocial punishment exceed 0.27. Parameters were fixed at $r=0.01$, $\gamma=0.1$, $\beta=0.3$, and $\rho=0$. The results shown were obtained under a scenario using the only "copy the successful" rule (PW-Fermi rule).}
    \label{figs1}
\end{figure*}

\vfill
\clearpage
\phantom{Invisible text.}
\vfill
\begin{figure*}[!h]
    \centering
\includegraphics[width=0.58\linewidth]{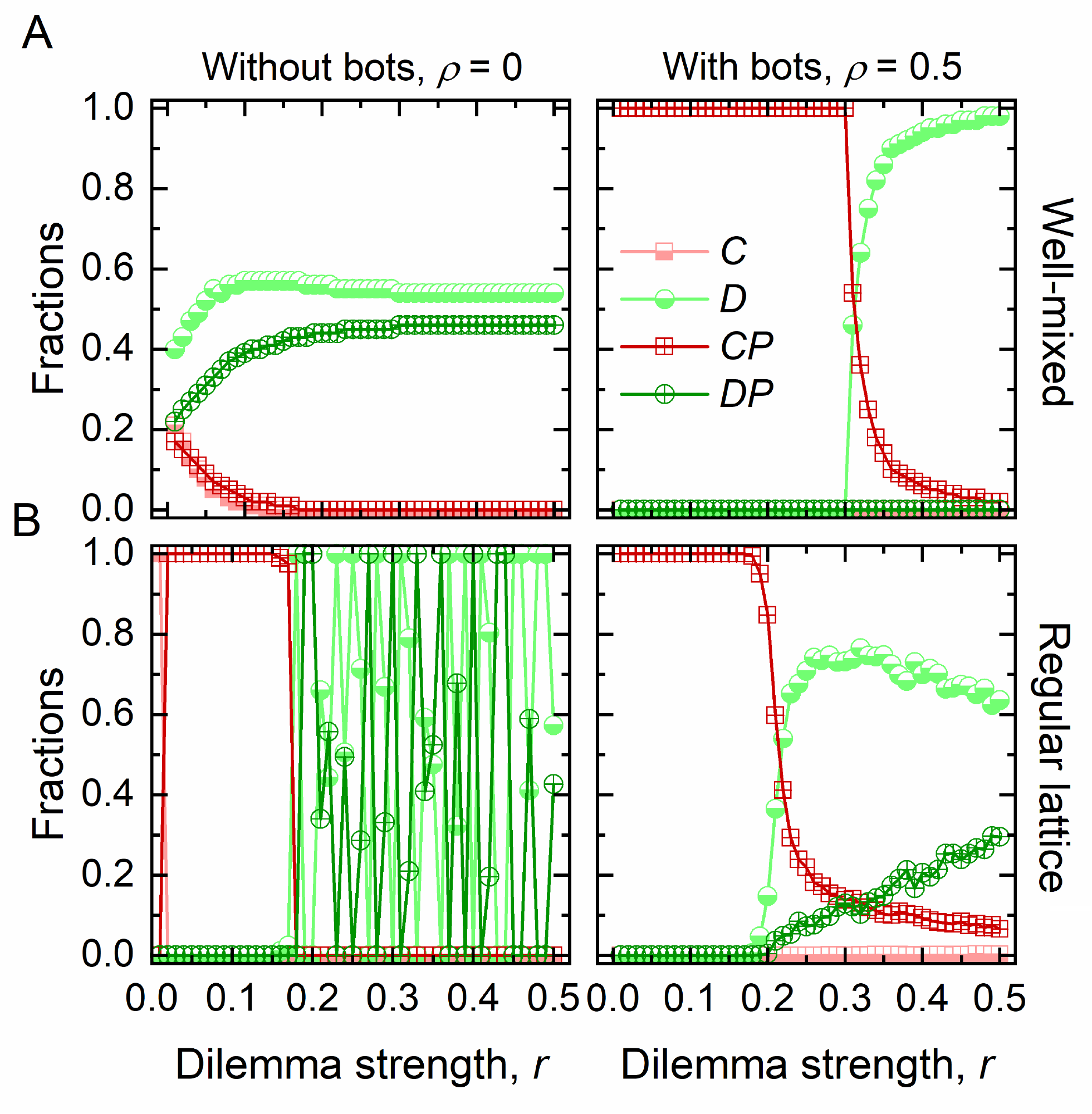}
    \caption{\textbf{Simple bots expand the survival region of prosocial punishment among humans.} Shown are the fractions of each actor as a function of dilemma strength. The top and bottom rows show the results of an infinite and well-mixed population and a regular lattice, respectively. The left and right columns represent scenarios without bots and with bots, respectively. The results shown were obtained under a scenario using the only "copy the successful" rule (PW-Fermi rule).}
    \label{figs2}
\end{figure*}
\vfill
\clearpage
\phantom{Invisible text.}
\vfill

\begin{figure*}[!h]
    \centering
\includegraphics[width=0.78\linewidth]{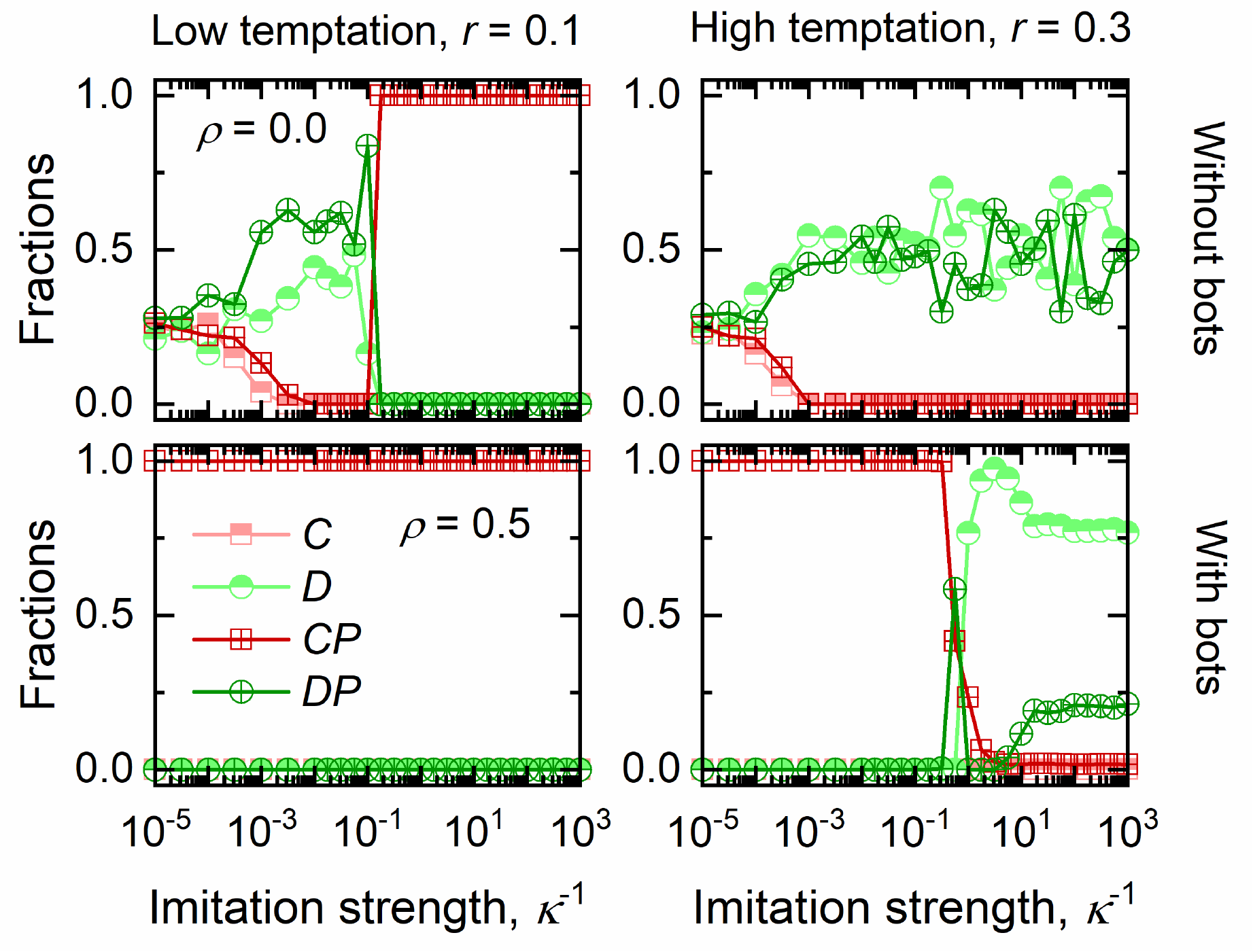}
    \caption{\textbf{The promotion effects of bots on prosocial punishment are only limited to weak imitation strength, while strong imitation strength diminishes the dominance of prosocial punishment.} Shown are the fractions of each actor as a function of imitation strength $\kappa^{-1}$. The top and bottom rows represent scenarios without and with bots, respectively. The left and right columns show the results of weak and strong dilemma strength, respectively. Parameters were fixed at $r=0.1$ (left column), $r=0.3$ (right column), $\rho=0$ (top row), and $\rho=0.5$ (bottom row). The results shown were obtained under a scenario using the only "copy t he successful" rule (PW-Fermi rule).}
    \label{figs2.1}
\end{figure*}
\vfill
\clearpage
\phantom{Invisible text.}
\vfill

\begin{figure*}[!h]
    \centering
    \includegraphics[width=0.86\linewidth]{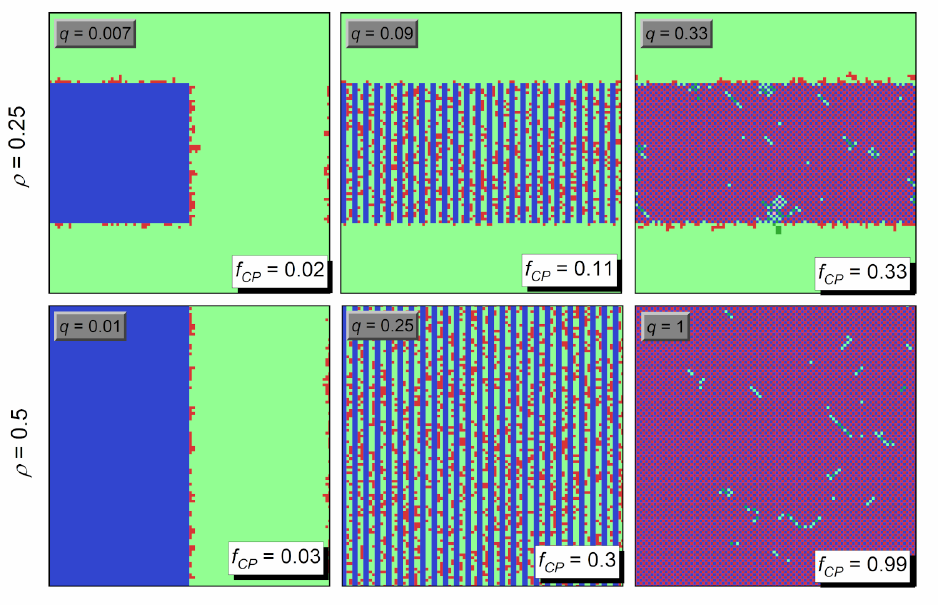}
      \caption{\textbf{Large interaction probabilities between bots and humans significantly improves the prosocial punishment level among humans.} Shown are stationary snapshots under different interaction probabilities $q$ between bots and humans. From left to right, the density of bots is unchanged. However, we varied the value of $q$ via different bot spatial configurations. Bots were assigned to the network in three different ways: (i) bots were assigned to the left domain of the network (left), (ii) bots were assigned to the network in stripes (middle), and (iii) bots were assigned to the network in a checkerboard pattern (right). By fixing the spatial configuration of the bots, we can easily calculate $q$, the interaction probability between bots and humans. The density of bots $\rho$ was fixed at 0.25 and 0.5 for the top and bottom rows, respectively. Bots are denoted in blue. Traditional cooperators and defectors, prosocial punishers, and antisocial punishers are depicted as light red, dark red, light green and dark green, respectively. It is clear that although the bot density remained unchanged, a large $q$ value leads to an optimal level of prosocial punishment among humans. Dilemma strength was fixed at $r=0.25$. The results shown were obtained under a scenario using the only "copy the successful" rule (PW-Fermi rule).}
    \label{figs3}
\end{figure*}

\vfill
\clearpage
\phantom{Invisible text.}
\vfill
\begin{figure*}[!h]
    \centering
\includegraphics[width=0.86\linewidth]{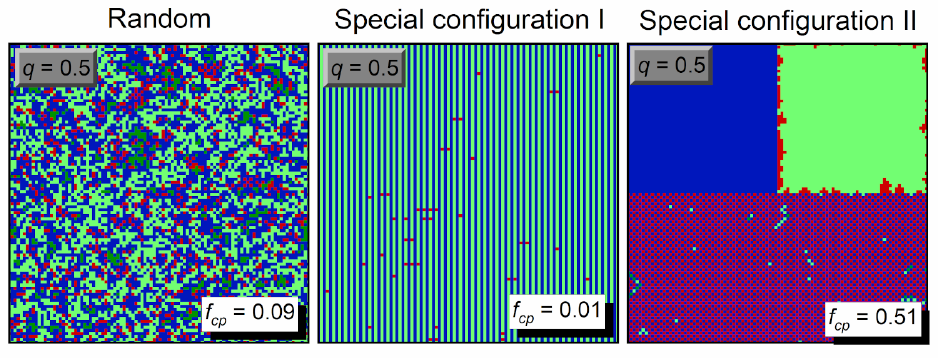}
    \caption{\textbf{In a regular lattice-structured population, bot spatial configuration is a key determinant for the prevalence of prosocial punishment.} The snapshots shown here were obtained under three different bot spatial configurations: (i), bots were randomly assigned to the network (left panel), (ii) bots were distributed to the network in stripes (middle panel), and (iii) 25\% of bots were assigned to half the domain of the network in a checkerboard fashion, and the other 25\% of the bots were assigned to the upper left region of the network (right panel). Bots are depicted in blue, $C$, $D$, $CP$, and $DP$ are depicted in light red, light green, dark red, and dark green, respectively. Parameter $q$ denotes the probability that normal players interact with bots, while $f_{cp}$ denotes the final prosocial punishment level of the population. The fraction of bots was fixed at 0.5 for all panels. It is clear that although the interaction probability between bots and normal players are equal in different scenarios, the bot spatial configuration matters and the chessboard configuration generates a better outcome. The dilemma strength $r$ was fixed at 0.25. The results shown were obtained under a scenario using the only "copy the successful" rule (PW-Fermi rule).}
    \label{fig3}
\end{figure*}

\vfill
\clearpage
\phantom{Invisible text.}
\vfill

\begin{figure*}[!h]
    \centering
    \includegraphics[width=0.86\linewidth]{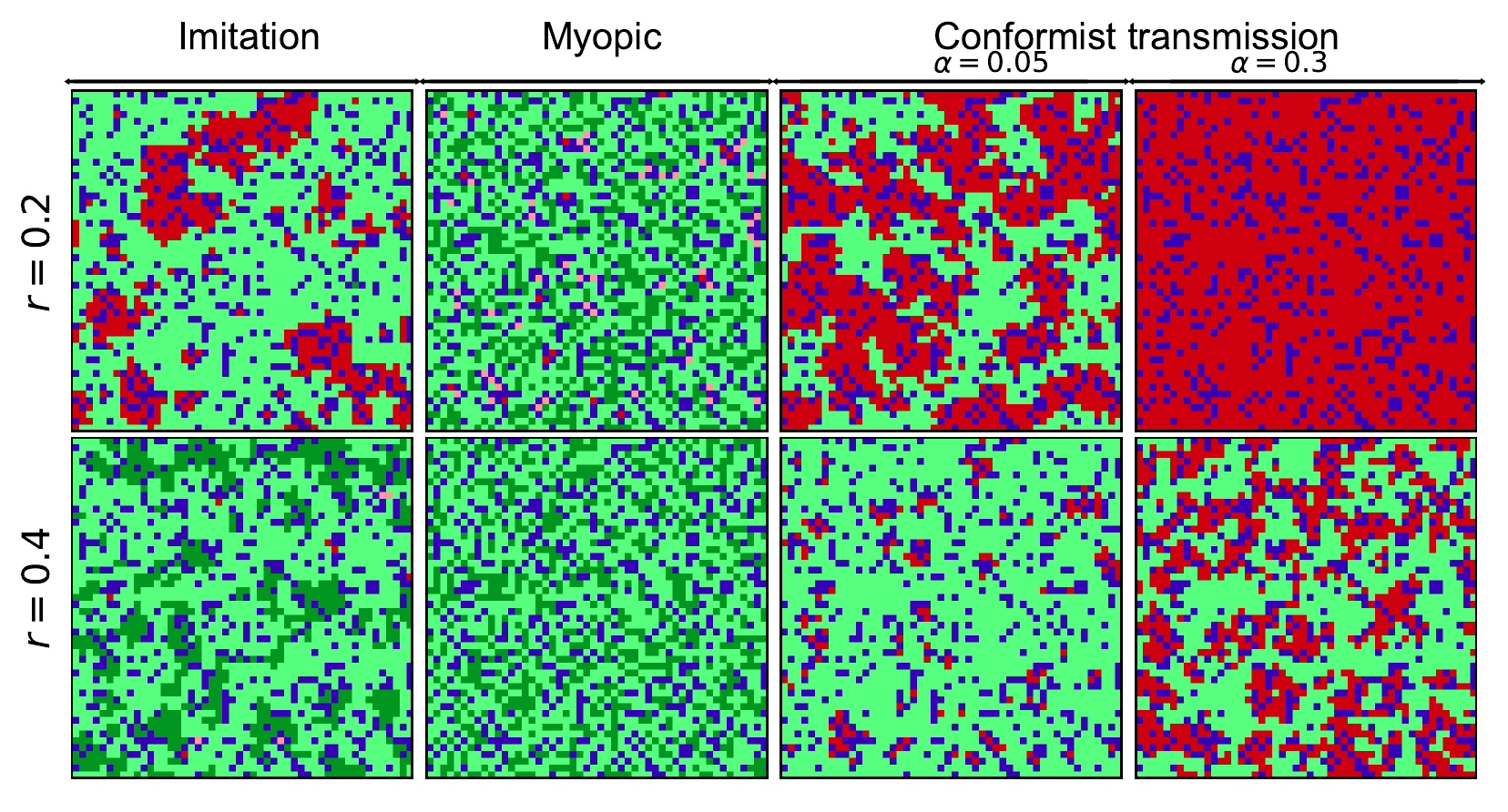}
     \caption{\textbf{Bots establish prosocial punishment among humans. Moreover, if normal players have a learning bias toward a "copy the majority" rule, the prosocial punishment level can be significantly improved in a regular lattice.} Shown are snapshots at a stationary state for four situations: scenarios implementing only a "copy the successful rule (PW-Fermi rule in the left panel), only a "myopic best response" rule (second column), weak conformist transmission (third column), and strong conformist transmission (right column). We used $r=0.2$ and $r=0.4$ for the top and bottom panels, respectively. The fraction of bots was fixed at 0.2. The conformist transmission rates were fixed at: $\alpha=0.05$ (third column) and $\alpha=0.3$ (right column) for the weak and strong conformist transmission scenarios, respectively. It is clear that bots can attract only a few prosocial punishers if humans care only about their material gain (i.e., they follow either the "copy the successful" rule or the "myopic best response" rules). However, if humans have a psychological bias toward a "copy the majority" rule, then prosocial punishment is significantly improved under weak conformist transmission and can even dominate the population under strong conformist transmission. All bots were predesigned to choose a $CP$ action. Bots are depicted in blue. $C$, $D$, $CP$, and $DP$ are depicted in light red, light green, dark red, and dark green, respectively.}
    \label{figs4}
\end{figure*}

\vfill
\clearpage
\phantom{Invisible text.}
\vfill

\begin{figure*}[!h]
    \centering
\includegraphics[width=0.91\linewidth]{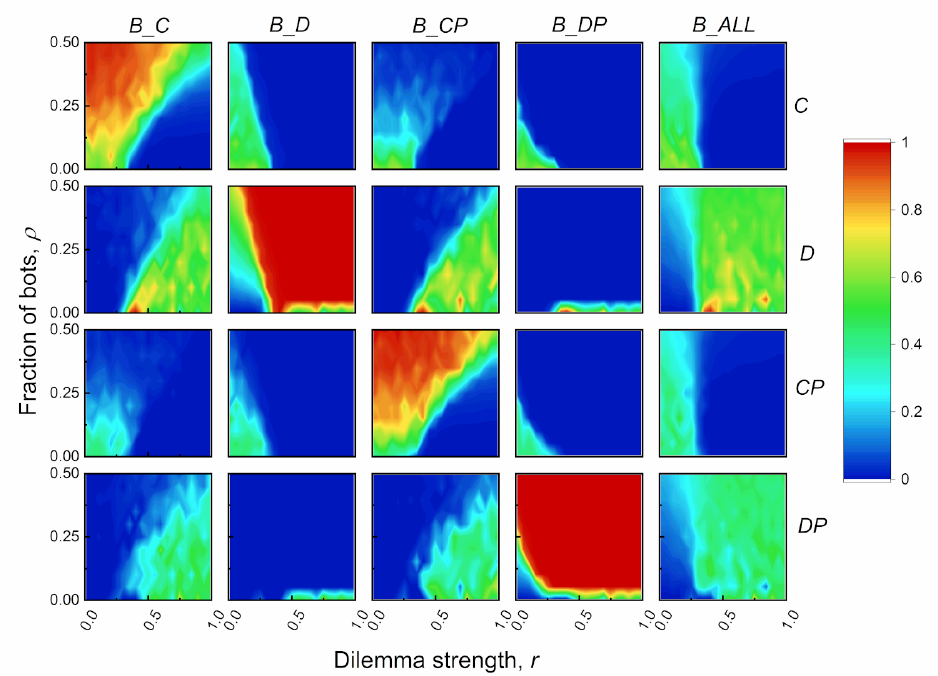}
    \caption{\textbf{In scale-free networks, cooperative bots produce significant improvements for cooperative actors, while defective bots diminish the cooperation-promoting effect. }  Shown are the fractions of each actor at a stationary state as a function of dilemma strength $r$ and bot density $\rho$ under five situations: (i) bot action was predesigned as: $C$ (marked as $B\_C$ in the first column); (ii) bot action was predesigned as: $D$ (marked as $B\_D$ in the second Column), (iii) bot action was predesigned as: $CP$ (marked as $B\_{CP}$ in the third column), (iv) bot action was predesigned as: $DP$ (marked by $B\_DP$ in the fourth column), and (v) a combination of all four aforementioned bots ($B\_C$, $B\_D$, $B\_CP$, and $B\_DP$) were introduced into the population with equal probability $\rho/4$ (marked as $B\_ALL$ in the last column).  From top to bottom, results correspond to the fractions of actions $C$, $D$, $CP$, $DP$ at a stationary state, and the average social payoff ($ASP$), respectively.  All results shown were obtained using only the pairwise Fermi imitation rule.} 
    \label{figs5}
\end{figure*}

\vfill
\clearpage
\phantom{Invisible text.}
\vfill

\begin{figure*}[!h]
    \centering
\includegraphics[width=0.91\linewidth]{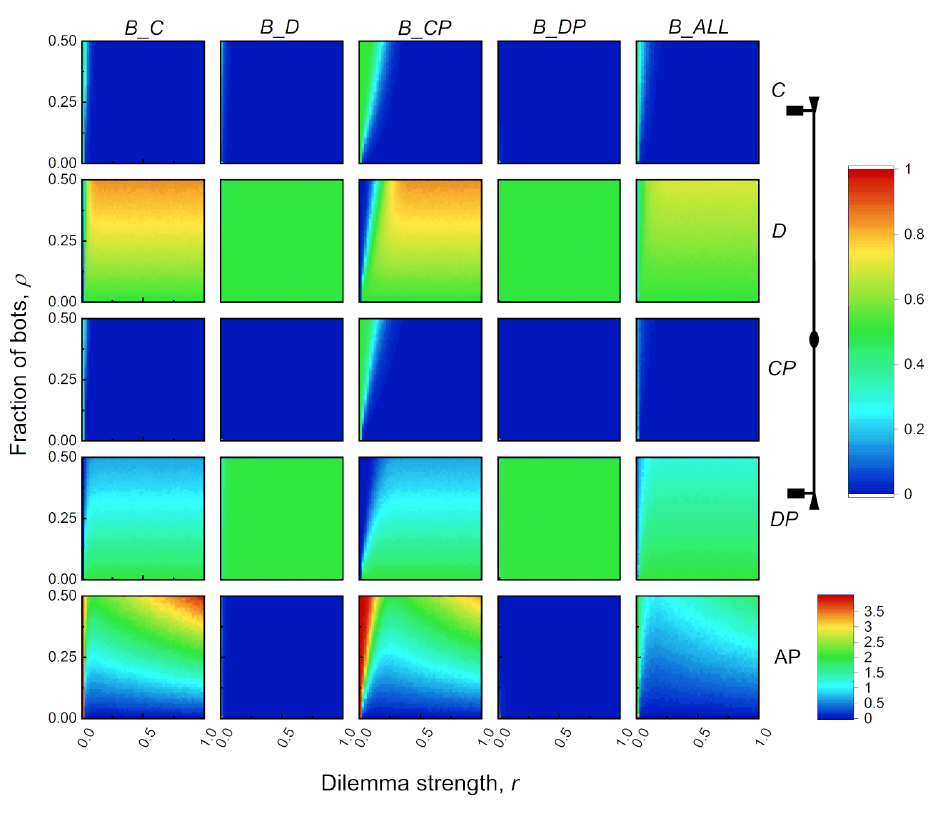}
    \caption{\textbf{The myopic best response rule produces similar outcomes in either a well-mixed population or a regular lattice.}  Shown are the fractions of each actor at a stationary state as a function of dilemma strength $r$ and bot density $\rho$ under five situations: (i) bot action was predesigned as: $C$ (marked as $B\_C$ in the first column); (ii) bot action was predesigned as: $D$ (marked as $B\_D$ in the second Column), (iii) bot action was predesigned as: $CP$ (marked as $B\_{CP}$ in the third column), (iv) bot action was predesigned as: $DP$ (marked by $B\_DP$ in the fourth column), and (v) a combination of all four aforementioned bots ($B\_C$, $B\_D$, $B\_CP$, and $B\_DP$) were introduced into the population with equal probability $\rho/4$ (marked as $B\_ALL$ in the last column).  From top to bottom, results correspond to the fractions of actions $C$, $D$, $CP$, $DP$ at a stationary state, and the average social payoff ($AP$), respectively. All results shown were obtained under a scenario using the only "myopic best response" rule.}
    \label{figs6}
\end{figure*}

\vfill
\clearpage
\phantom{Invisible text.}
\vfill

\begin{figure*}[!h]
    \centering
\includegraphics[width=0.91\linewidth]{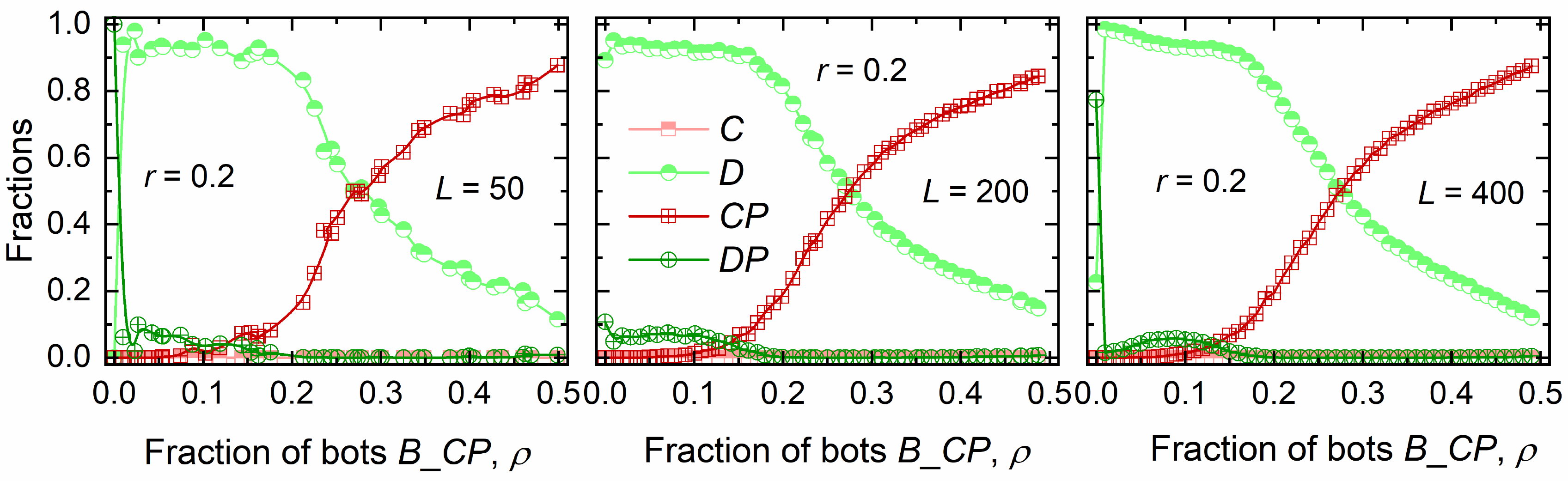}
    \caption{\textbf{The effect of the bots in encouraging prosocial punishment is robust against network size.} Shown are the fractions of each actor at a stationary state as a function of bot density, $\rho$. From left to right the employed network size is $50 \times 50$, $200\times200$, and $400\times400$, respectively. Dilemma strength was fixed at: $r=0.5$. We used the solely the PW-Fermi rule and the value of $\alpha$ was fixed at 0. Results were obtained for the bots predesigned to choose $CP$. }
    \label{figs7}
\end{figure*}

\vfill
\clearpage
\phantom{Invisible text.}
\vfill

\begin{figure*}[!h]
    \centering
\includegraphics[width=0.91\linewidth]{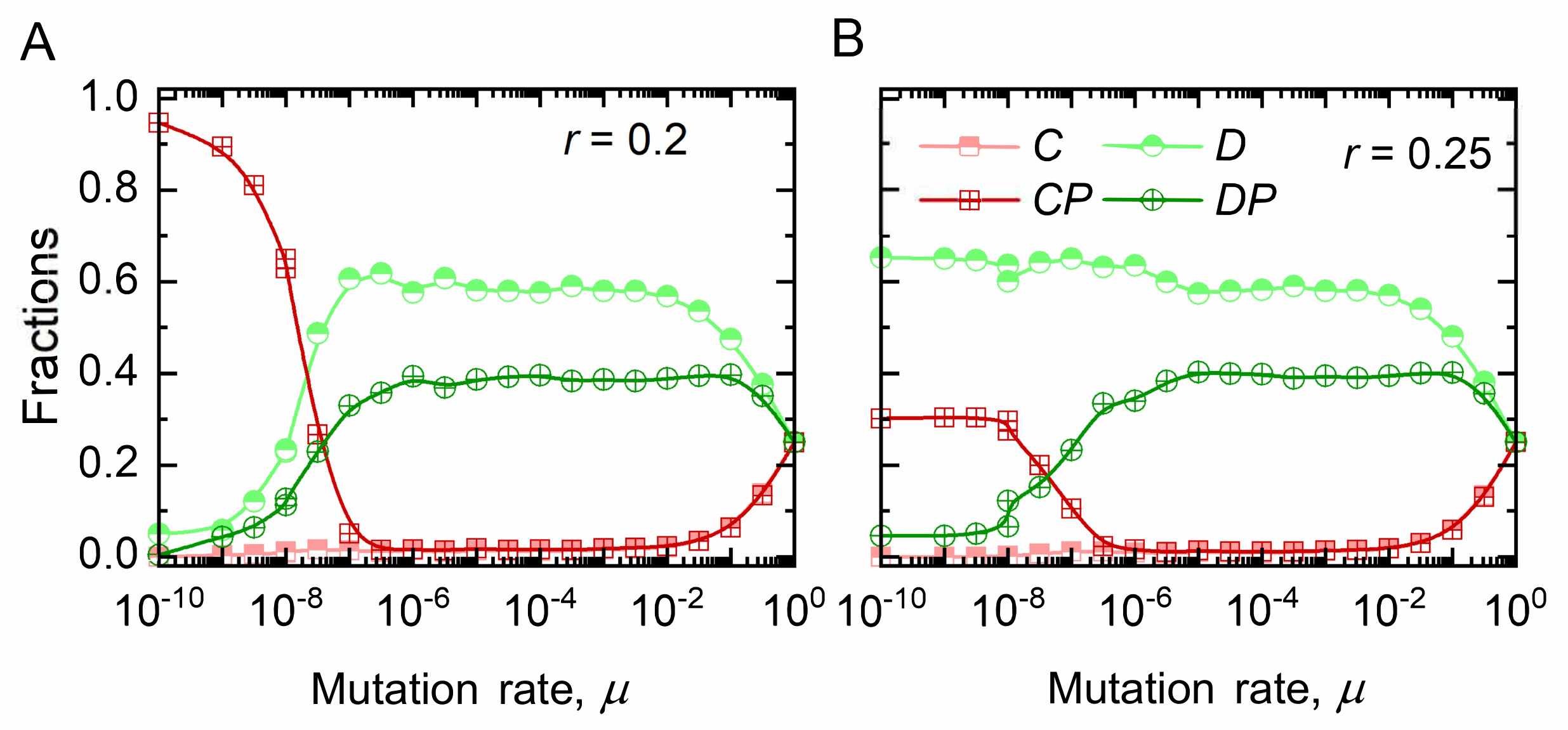}
    \caption{\textbf{The effect of bots on humans are robust against small mutation rate, while defective actors dominate the whole population with increasing mutation rate.} Shown are the fractions of each actor independence on mutation rate. We fixed the dilemma strength as 0.2 and 0.25 for left and right column, respectively. The density of bots was fixed as 0.5. Results were obtained for the bots with $CP$, we used the solely the PW-Fermi rule and the value of $\alpha$ was fixed as 0.}
    \label{figs8}
\end{figure*}

\vfill
\clearpage
\phantom{Invisible text.}
\vfill

\begin{figure*}[!h]
    \centering
\includegraphics[width=0.88\linewidth]{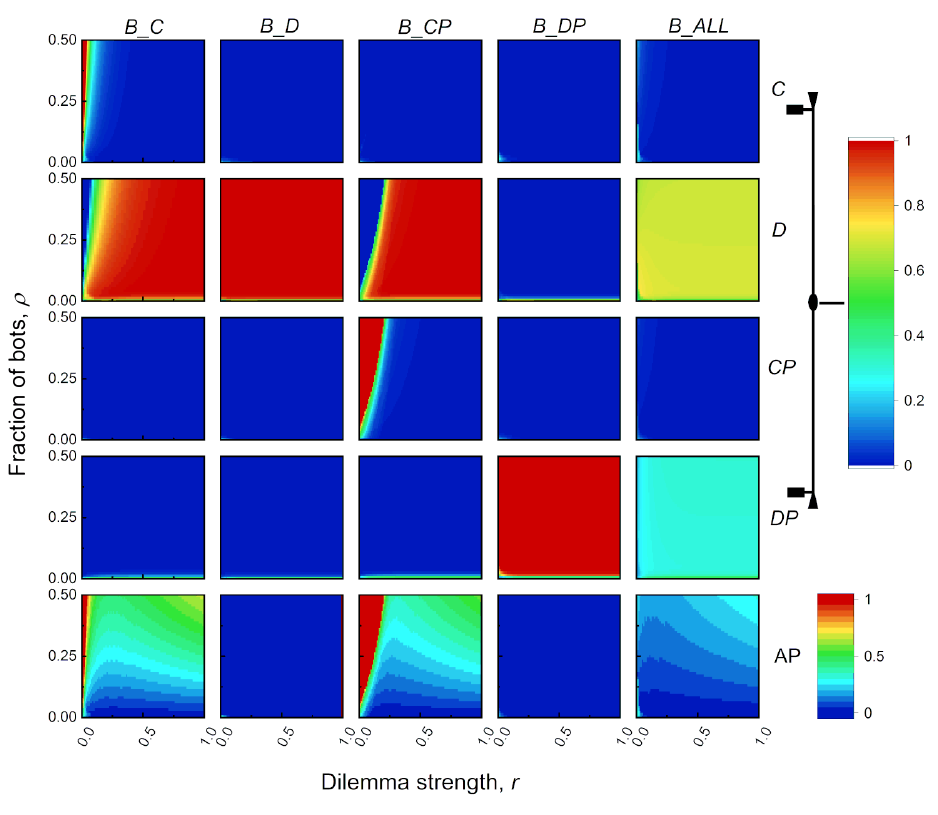}
     \caption{\textbf{In a well-mixed population, cooperative bots breed and promote cooperative actors, but bots designed to choose $CP$ produce much more significant improvement in cooperation than bots choosing $C$. } Shown are the fractions of each actor at a stationary state as a function of dilemma strength $r$ and bot density $\rho$ under five situations: (i) bot action was predesigned as: $C$ (marked as $B\_C$ in the first column); (ii) bot action was predesigned as: $D$ (marked as $B\_D$ in the second Column), (iii) bot action was predesigned as: $CP$ (marked as $B\_{CP}$ in the third column), (iv) bot action was predesigned as: $DP$ (marked by $B\_DP$ in the fourth column), and (v) a combination of all four aforementioned bots ($B\_C$, $B\_D$, $B\_CP$, and $B\_DP$) were introduced into the population with equal probability $\rho/4$ (marked as $B\_ALL$ in the last column).  From top to bottom, results correspond to the fractions of actions $C$, $D$, $CP$, $DP$ at a stationary state, and the average social payoff ($AP$), respectively.  All results shown were obtained using only the pairwise Fermi imitation rule.}
    \label{figs9}
\end{figure*}
\vfill
\clearpage
\phantom{Invisible text.}
\vfill

\begin{figure*}[!h]
    \centering
\includegraphics[width=0.88\linewidth]{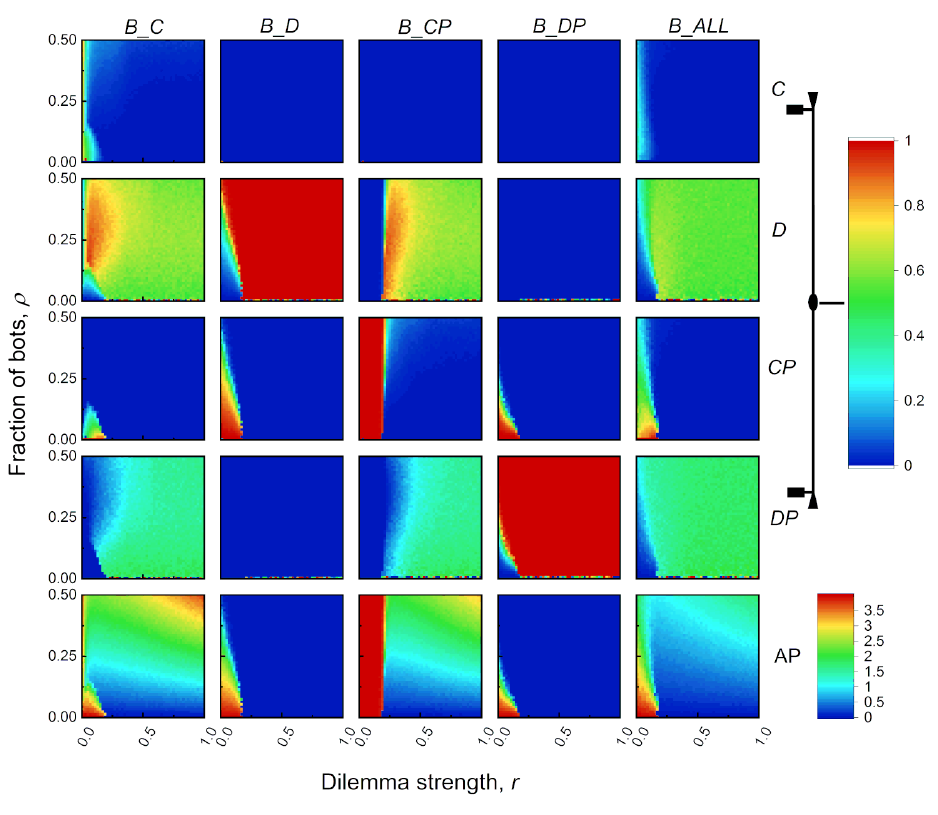}
    \caption{\textbf{In regular lattice, increasing the density of the bots predesigned to choose $CP$ improves the prosocial punishment level among normal players, while increasing the density of the bots predesigned to choose $C$ harms cooperation.} Shown are the fractions of each actor at a stationary state as a function of dilemma strength $r$ and bot density $\rho$ under five situations: (i) bot action was predesigned as: $C$ (marked as $B\_C$ in the first column); (ii) bot action was predesigned as: $D$ (marked as $B\_D$ in the second Column), (iii) bot action was predesigned as: $CP$ (marked as $B\_{CP}$ in the third column), (iv) bot action was predesigned as: $DP$ (marked by $B\_DP$ in the fourth column), and (v) a combination of all four aforementioned bots ($B\_C$, $B\_D$, $B\_CP$, and $B\_DP$) were introduced into the population with equal probability $\rho/4$ (marked as $B\_ALL$ in the last column).  From top to bottom, results correspond to the fractions of actions $C$, $D$, $CP$, $DP$ at a stationary state, and the average social payoff ($AP$), respectively.  All results shown were obtained using only the pairwise Fermi imitation rule, and the value of $\alpha$ was fixed at 0.}
    \label{figs10}
\end{figure*}

\vfill
\clearpage
\phantom{Invisible text.}
\vfill

\begin{figure*}[!h]
    \centering
\includegraphics[width=0.91\linewidth]{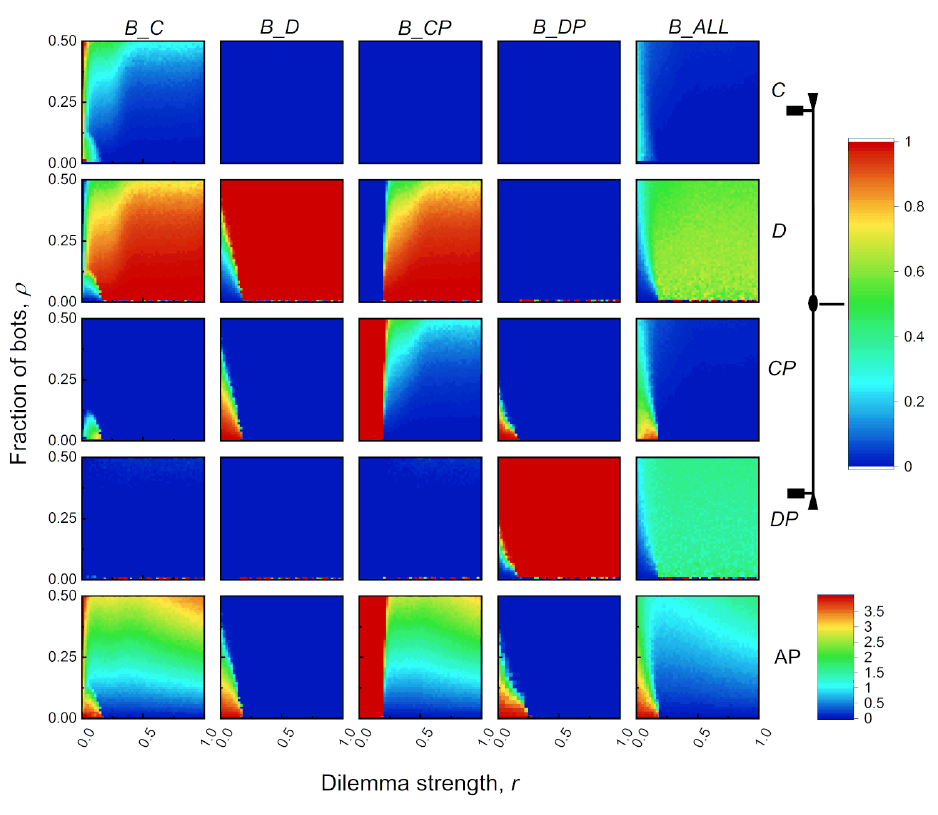}
    \caption{\textbf{Weak conformist transmission generates results similar to figure.\ref{figs10}, but produces significant improvement in prosocial punishment.} The parameters and markers are the same as for figure.\ref{figs10}. The conformist transmission rate was fixed at 0.05.}
    \label{figs11}
\end{figure*}

\vfill
\clearpage
\phantom{Invisible text.}
\vfill

\begin{figure*}[!h]
    \centering
\includegraphics[width=0.91\linewidth]{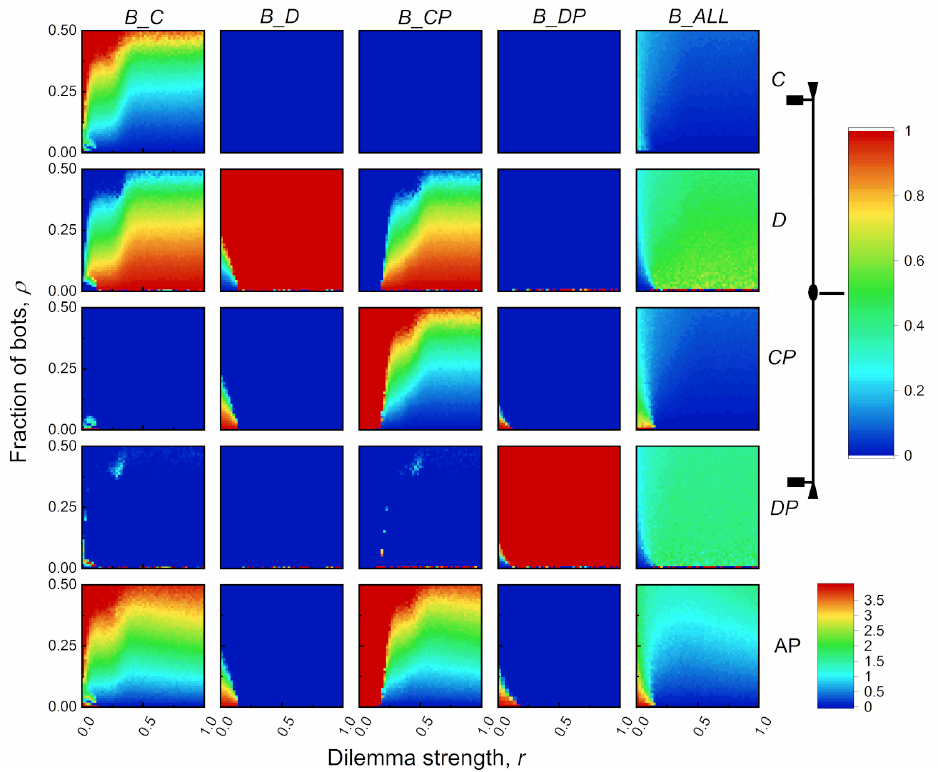}
    \caption{\textbf{Strong conformist transmission produces more significant improvement in prosocial punishment than weak conformist transmission.} The parameters and markers are the same as for figure.\ref{figs10}. The conformist transmission rate was fixed at 0.3.}
    \label{figs12}
\end{figure*}

\end{document}